\documentclass[journal]{IEEEtran}
\usepackage{subfigure}
\usepackage{colortbl}
\usepackage{bm}

\usepackage{graphicx}  
\usepackage{url}       

\usepackage{amsmath}   
\allowdisplaybreaks[4]
\usepackage{cite}

\usepackage{amsfonts,amssymb}

\usepackage{stfloats}

\usepackage{cases}
\usepackage{algorithm}
\usepackage{multirow}
\usepackage{algorithmic}
\usepackage{epstopdf}
\usepackage{color}

\usepackage[square, comma, sort&compress, numbers]{natbib}

\makeatletter
\newcommand{\vvast}{\bBigg@{3.0}}
\newcommand{\vast}{\bBigg@{4}}
\newcommand{\Vast}{\bBigg@{4.5}}
\newcommand{\VVast}{\bBigg@{5}}
\newcommand{\VVVast}{\bBigg@{5.5}}

\makeatother

\newcommand{\ls}[1]
{\dimen0=\fontdimen6\the\font
	\lineskip=#1\dimen0
	\advance\lineskip.5\fontdimen5\the\font
	\advance\lineskip-\dimen0
	\lineskiplimit=.9\lineskip
	\baselineskip=\lineskip
	\advance\baselineskip\dimen0
	\normallineskip\lineskip
	\normallineskiplimit\lineskiplimit
	\normalbaselineskip\baselineskip
	\ignorespaces
}

\begin{document}

	
		\title{\ls{1.0}E2-Conditioned Finite-Horizon Effective Capacity for Public-Safety MCX over Shared O-RAN}

\author{\IEEEauthorblockN{Jingqing Wang,~\IEEEmembership{Member,~IEEE} and Wenchi Cheng,~\IEEEmembership{Senior Member,~IEEE}} \\[0.2cm]
	
	\thanks{
This work of Jingqing Wang and Wenchi Cheng was supported in part by the National Natural Science Foundation of China under Grant 62341132 and the Natural Science Basic Research Program of Shaanxi under Grant 2024JC-YBQN-0642.}
	\thanks{Jingqing Wang and Wenchi Cheng are with the State Key Laboratory of Integrated Services Networks, Xidian University, Xi'an, China (e-mails: jqwangxd@xidian.edu.cn; wccheng@xidian.edu.cn).}
}

\maketitle


\begin{abstract}
Supporting public-safety Mission Critical Services (MCX) over a shared Open radio access network (O-RAN) requires service assurance over finite incident horizons, while ordinary mobile traffic competes for the same resources and heterogeneous E2 domains expose different observations, control actions, and actuation latencies. 
Existing RAN key performance indicators are retrospective, whereas conventional effective capacity characterizes an asymptotic stationary regime and therefore suppresses both the initial E2-observed condition and the finite selection-to-actuation transient. 
To solve this problem, in this paper we develop an E2-conditioned finite-horizon effective capacity (FH-EC) framework for public-safety MCX over shared O-RAN. 
Specifically, we first establish a finite-horizon O-RAN MCX-based service model that incorporates E2-observed network states, control actuation latency, and correlation-aware connectivity diversity. Based on this model, we derive an FH-EC formulation that characterizes the executable service capability within a finite mission horizon.
Furthermore, we transform FH-EC into confidence-calibrated capability profiles and develop an MCX orchestration framework with contract certification, shared-resource protection, and FH-EC-driven profile selection at the Near-RT RAN Intelligent Controller (RIC).
Coupled MATLAB/ns-3 evaluations demonstrate the predicted short-horizon state and actuation effects and show that adaptive connectivity selection improves MCX supportability under O-DU degradation while satisfying the configured multi-QoS and non-MCX protection requirements.


\end{abstract}

\begin{IEEEkeywords}
Finite-horizon effective capacity, multi-QoS provisioning, E2 exposure, HRLLC, public-safety MCX, O-RAN.
\end{IEEEkeywords}

\section{Introduction}\label{sec:intro}

\IEEEPARstart{O}{pen} radio access network (O-RAN) has emerged as a promising architecture for next-generation networks by introducing programmable control, open interfaces, and AI-driven network orchestration~\cite{10969847}.  
Through the Service Management and Orchestration (SMO) framework, the Non-Real-Time RAN Intelligent Controller (Non-RT RIC), the Near-Real-Time RIC (Near-RT RIC), and standardized interfaces, network intelligence can be deployed across disaggregated O-RAN Central Unit (O-CU) and O-RAN Distributed Unit (O-DU) entities operating under heterogeneous implementations and administrative domains~\cite{10772472}. 
This architectural evolution creates an important opportunity for supporting public-safety Mission Critical Services (MCX) over shared mobile infrastructure~\cite{3GPP_TS280,11169488}. However, programmability alone does not provide service assurance. Open interfaces make a RAN configuration observable and controllable, but not necessarily certifiably supportable for a critical service. 

This distinction is particularly important for MCX. Control and group-voice bearers require bounded delivery delay and a low probability of terminal non-delivery. Real-time video is additionally sensitive to packet-delay variation while status-update traffic requires information freshness.
Finite-blocklength transmission further couples payload rate, blocklength, and reliability~\cite{5452208}. 
Consequently, delivered rate, delay, jitter, age of information (AoI)~\cite{9380899,10547087}, and reliability represent different, service-dependent quality-of-service (QoS) dimensions rather than interchangeable manifestations of average throughput~\cite{10609803,9635675,11551157}. Their interactions are central to hyper-reliable and hyper-low-latency communications (HRLLC)~\cite{ITU01,11480692,11456336}. 
In a shared RAN, these requirements are coupled to ordinary mobile services. Reserving additional resources may improve MCX supportability, but can substantially reduce non-MCX utility or starve a protected subset of ordinary users. 
An MCX assurance mechanism must therefore certify service-specific risks while explicitly constraining the collateral degradation imposed on non-MCX traffic.

Although O-RAN provides the architectural mechanisms needed to act on this problem, it introduces a capability-semantic gap. E2 measurements describe what an E2 node reports, while its advertised RAN functions specify what the Near-RT RIC may observe and control. Different E2 nodes may expose different measurement granularity, reporting periods, control actions, and actuation delays. 
These raw observations and action sets do not directly quantify the service that an executable configuration can support. At the admission epoch of an MCX flow, the Near-RT RIC must select a profile specifying the connectivity mode, participating domains, local execution policy, and reserved resources. 
The central challenge, therefore, is to determine the level of service a candidate profile can guarantee within a finite MCX mission window~\cite{10221713}, accounting for real-time E2 telemetry and selection-actuation delays. Addressing this requires a unified capability coordinate system through which heterogeneous E2 configurations can be compared and safely selected.

The mission window in this question is not merely the periodic execution interval of the Near-RT RIC. It is the finite commitment window over which an admitted MCX profile and its service certificate must remain valid. 
After admission, slot-level scheduling, hybrid automatic repeat request (HARQ), link adaptation, and other local decisions may continue to adapt within the selected O-CU/O-DU policy, but the certified profile cannot be freely replaced without invalidating the capability. 
Moreover, profile-selection runtime, E2 control-message delivery, processing, and distributed O-CU/O-DU actuation consume a non-negligible portion of a short mission window. During this transient, service is governed by the pre-actuation policy rather than the newly selected one. The service available over the complete mission window therefore depends jointly on the E2-conditioned initial state, the finite control transient, and the post-actuation execution policy.

Existing RAN key performance indicators cannot directly answer this finite-horizon question because they are primarily descriptive rather than forward-looking and action-conditioned. 
From a theoretical perspective, effective capacity provides a risk-sensitive abstraction for statistical delay provisioning~\cite{1210731,10879302}. 
However, conventional effective capacity is defined through the asymptotic log-moment rate of a stationary and ergodic service process. Its steady-state limit suppresses dependence on the initial E2-observed condition and assigns a vanishing contribution to a finite control transient. 
This approximation is appropriate when the operational horizon is much longer than the channel mixing time, congestion-burst duration, mobility interval, infrastructure-recovery time, and control-actuation delay. It can, however, overestimate or underestimate the service available when the MCX commitment window is comparable to these timescales. The required abstraction should therefore retain the risk-sensitive interpretation of effective capacity over the actual mission horizon, explicitly incorporate the selection-and-actuation transient, and recover conventional effective capacity as the horizon grows.

Existing studies provide complementary components of such an abstraction. Finite-horizon scheduling has been investigated for wireless networked control systems~\cite{10221713}, while finite-blocklength and mission-effective-capacity formulations characterize communication reliability and dependability~\cite{9512400}. 
In O-RAN, resource allocation has been studied under network-slicing constraints~\cite{9888767}, and stochastic network calculus has been used to support latency-tail-sensitive orchestration~\cite{10621356}. ColO-RAN and ns-O-RAN further provide valuable foundations for learning-based xApps, closed-loop control, and system-level O-RAN experimentation \cite{9814869,lacava2023ns}. Packet duplication and multi-connectivity can improve delivery reliability by exploiting spatial diversity, although their benefit depends on resource consumption and correlated radio or infrastructure failures \cite{8329621}. These studies establish the foundations for programmable control, statistical QoS provisioning, and connectivity diversity. 
However, they do not jointly provide an E2-conditioned and actuation-aware finite-horizon capability contract that carries simultaneous finite-sample service and multi-QoS certificates and can be consumed by a Near-RT RIC while protecting ordinary traffic in a shared O-RAN.

To bridge this gap, we develop an E2-conditioned finite-horizon effective-capacity (FH-EC) framework for public-safety MCX over shared O-RAN. 
Specifically, first, we develop an E2-conditioned and actuation-aware finite-horizon service model for MCX over shared O-RAN. 
Second, we establish the long-horizon consistency of the proposed FH-EC under a finite E2-control transient and a primitive post-actuation Markov-additive kernel. 
Third, we develop a contract-preserving O-RAN profile-orchestration framework. 
Coupled MATLAB/ns-3 system-level evaluation results evaluate the capability model and assess the effects of the proposed E2-conditioned finite-horizon multi-QoS orchestration.

The rest of this paper is organized as follows. Section~II builds the O-RAN MCX service model. Section~III derives E2-conditioned FH-EC expressions. Section~IV establishes the FH-EC-driven MCX profile orchestration in O-RAN. Section V validates and assesses the system performance. Section VI concludes the paper.

	\section{O-RAN MCX Service Model}\label{sec:sys}

\begin{figure}[!t]
	\centering
	\includegraphics[scale=0.46]{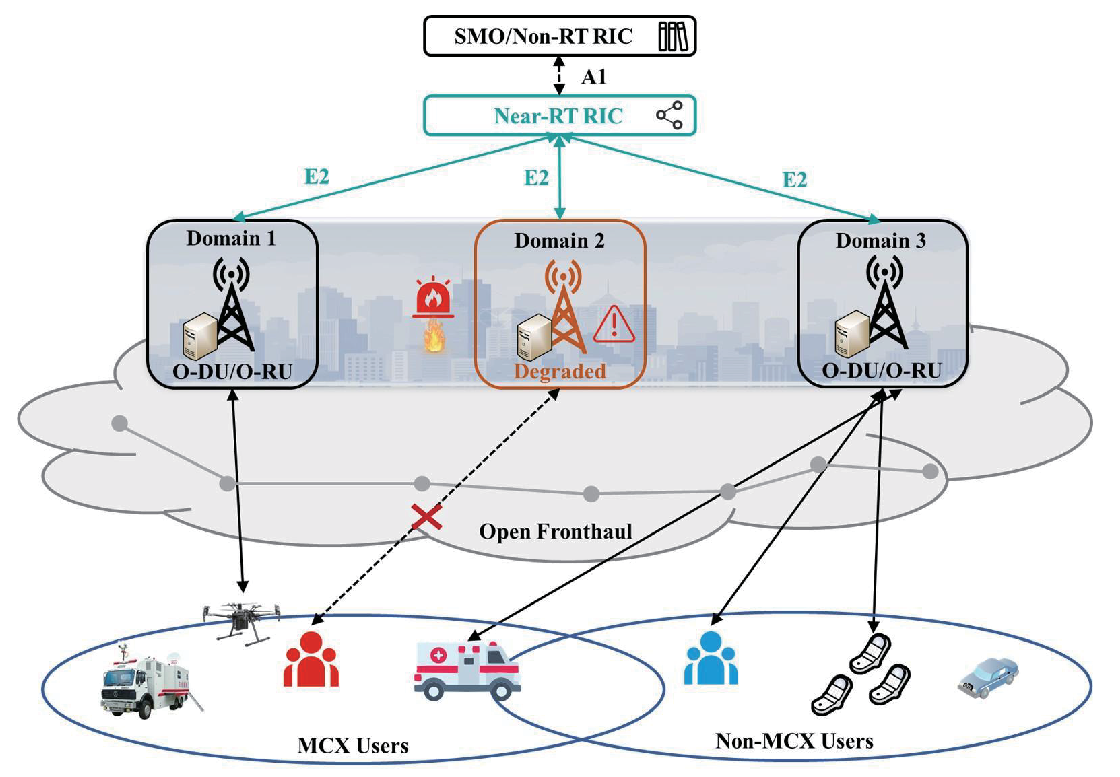}
	\caption{Shared Open 6G RAN architecture for E2-conditioned FH-EC service profile orchestration. }
	\label{fig1}
\end{figure}

\subsection{Network Architecture and Information Model}
As illustrated in Fig.~\ref{fig1}, the considered O-RAN architecture comprises an SMO/Non-RT RIC, a logical Near-Real-Time RIC coordination plane, one or more O-CU-CP/O-CU-UP entities, and multiple O-DU/O-RAN Radio Unit (O-RU) nodes. 
The O-CU-UP hosts the user-plane Packet Data Convergence Protocol (PDCP) functions, while Radio link Control (RLC), MAC, HARQ, and slot-level scheduling are executed at the O-DUs~\cite{10024837,pan01}.

Let ${\cal I}=\{1,\dots, I\}$ be the logical E2 control domains.
Domain $i\in{\cal I}$ contains the O-DU/E2-node set, denoted by ${\cal B}_{i}$, where ${\cal B}=\cup_{i\in{\cal I}}{\cal B}_{i}$ and ${\cal B}_{i}\cap {\cal B}_{j}=\emptyset, i\neq j$, and  ${\cal B}$ is the set of O-DUs. 
Different E2 nodes may expose different measurements, control functions, and actuation delays. We characterize the E2 exposure profile, denoted by $\bm{\Xi}_{i}$, of domain $i$ by
\begin{equation}\label{equation00}
	\bm{\Xi}_{i}\triangleq \left(\bm{h}_{i}^{\text{KPM}},{\cal A}_{i}^{\text{RC}},T_{i}^{\text{KPM}},\tau_{i,l}^{\text{act}}\right)
\end{equation}
where $\bm{h}_{i}^{\text{KPM}}$ is the domain-specific Key Performance Measurement (KPM) observation and aggregation map, ${\cal A}_{i}^{\text{RC}}$ is the set of RAN-control commands exposed to the orchestrator, $T_{i}^{\text{KPM}}$ is the KPM reporting period, and $\tau_{i,l}^{\text{act}}$ is the delay from issuing a control command to its effective application.
Let $T_{sl}$ and $T_{x}$ denote the O-DU scheduling-slot duration and the Near-RT RIC orchestration interval, respectively with $T_{sl}\ll T_{x}$.
We define $t_{l}=lT_{x}$ as the beginning of Near-RT orchestration epoch $l$, and let $k_{l}=\lfloor t_{l}/T_{sl} \rfloor$ as the corresponding slot index. 
The observation delay, denoted by $\delta_{i,l}^{\text{obs}}$,  measured in slots is given as follows:
\begin{equation}
		\delta_{i,l}^{\text{obs}}=\lceil \frac{\tau_{i,l}^{\text{obs}}}{T_{sl}}\rceil.
\end{equation}
Let $\bm{z}_i[k]$ denote the physical state of domain $i$, and let $h_i^{\mathrm{KPM}}$ be its E2SM-KPM observation and aggregation map. Then, the latest E2-exposed observation, denoted by $\hat{\bm{z}}_{i}[l]$, available at $t_{l}$ is modeled as:
\begin{equation}
	\widehat{\bm{z}}_i[l]
	=
	h_i^{\mathrm{KPM}}\!\left(
	\bm{z}_i[k_l-\delta_{i,l}^{\mathrm{obs}}]
	\right)
	+
	\boldsymbol\nu_i[l]
\end{equation}
where $\boldsymbol\nu_i[l]$ represents measurement noise,
time/frequency aggregation error, quantization, compression,
and unreported-state uncertainty.
The information available to the orchestrator, denoted by ${\cal F}_{l}^{\text{E2}}$, consists of the time-stamped E2 observations and past exposed control outcomes.

An MCX flow $s$ is admitted at orchestration epoch $l_s$, whose
beginning time is $t_{l_s}=t_s^0$. At admission, the Near-RT RIC selects one persistent profile $a$, which is given as follows:
\begin{equation}
	a
	=
	\left(
	m_a,\mathcal I_a,\varpi_{s,a},
	\bm{b}_{s,a},\tau_{s,a}^{\mathrm{act}}
	\right),
	\label{eq:persistent_candidate}
\end{equation}
where $m_a\in\{\mathrm{SC},\mathrm{DP}\}$ is the connectivity mode,
$\mathcal I_a$ is the participating E2-domain set,
$\varpi_{s,a}$ is the fixed O-CU/O-DU execution-policy tuple,
$\bm{b}_{s,a}$ is the reserved-resource vector, and
$\tau_{s,a}^{\mathrm{act}}$ is the profile-actuation delay.
Once admitted, the profile identifier and its resource reservation remain unchanged over a finite operational window, denoted by ${\cal T}_{s}$,  given by
\begin{equation}\label{equation01}
	{\cal T}_{s}=[t_{s}^{0},t_{s}^{0}+\Delta_{s})
\end{equation}
where $\Delta_{s}$ is the mission horizon, which is given by 
\begin{equation}\label{equation02b}
	\Delta_{s}=K_{s}{\cal T}_{sl},  \qquad  K_{s}\in\mathbb{Z}_{+}
\end{equation}
Slot-level scheduling, HARQ, and link adaptation may still adapt to local observations according to the fixed local policy contained in $\varpi_{s,a}$. 

Let $T_{\mathrm{sel}}[l_s]$ denote the runtime required by the
Near-RT RIC to complete profile selection for flow $s$ at its
admission epoch $l_s$. The runtime is upper bounded as $0\leq T_{\mathrm{sel}}[l_s] \leq T_{\mathrm{sel}}^{\max}$. For candidate profile $a$, the total control-to-effect delay, denoted by $\tau_{s,a}^{\mathrm{ctl}}$, is defined as follows:
\begin{equation}
	\tau_{s,a}^{\mathrm{ctl}}
	\triangleq
	T_{\mathrm{sel}}^{\max}+\tau_{s,a}^{\mathrm{act}},
	\label{eq:total_control_delay}
\end{equation}
where $\tau_{s,a}^{\mathrm{act}}$ includes the E2 control-message
delivery, processing, and distributed O-CU/O-DU actuation delay.
The corresponding number of pre-actuation slots and the effective
slot of profile $a$ are, respectively, given as follows:
\begin{align}
	\begin{cases}
	d_{s,a}^{\mathrm{ctl}} \!\!\!\!\!\!
	&\triangleq
	\left\lceil
	\frac{\tau_{s,a}^{\mathrm{ctl}}}{T_{\mathrm{sl}}}
	\right\rceil; \\
	k_{s,a}^{\mathrm{eff}}\!\!\!\!\!\!
	&\triangleq
	k_{l_s}+d_{s,a}^{\mathrm{ctl}}.
		\end{cases}
	\label{eq:effective_slot}
\end{align}
Only profiles satisfying $d_{s,a}^{\mathrm{ctl}}<K_s$ are eligible for admission. During the interval
$[k_{l_s},k_{s,a}^{\mathrm{eff}})$, the target profile has not yet
become effective. We denote the corresponding pre-actuation policy
by $\varpi_s^{\mathrm{pre}}$. For a newly admitted MCX flow,
$\varpi_s^{\mathrm{pre}}$ provides no guaranteed payload service,
whereas for an explicitly supported reconfiguration it may represent
the incumbent policy. From slot $k_{s,a}^{\mathrm{eff}}$ onward,
the selected policy $\varpi_{s,a}$ remains unchanged until the end
of the mission window.

\subsection{Finite-Horizon SC/DP Radio Service}
Let ${\cal S}={\cal S}_{M}\cup {\cal S}_{N}$ and ${\cal S}_{M}\cap {\cal S}_{N}=\emptyset$, where ${\cal S}_{M}$ and ${\cal S}_{N}$ denote the MCX and non-MCX flow sets. An MCX bearer may represent a control stream, a video stream, or a data/status-update stream~\cite{3GPP_TS2802}.

Packet $n$ of flow $s$ arrives at the RAN ingress at time $g_{s,n}$ and contains $L_{s,n}$ information bits. The number of arriving bits, denoted by $A_{s}[k]$, in slot $k$ is given as follows:
\begin{equation}\label{equation03}
	A_{s}[k]=\sum_{n:g_{s,n}\in [kT_{sl},(k+1)T_{sl})} L_{s,n}.
\end{equation}


Let $\chi_{b}^{W}[k]\in[0,1]$ and $\chi_{b}^{P}[k]\in[0,1]$ denote the available bandwidth and power fractions of O-DU $b$. 
Let $\omega_{b,f}[k]$ and $p_{b,f}[k]$ denote the bandwidth and transmit power allocated by O-DU $b$ to flow $f\in{\cal S}$ in slot $k$. 
 The per-slot constraints are listed as follows:
\begin{equation}\label{equation04}
	\begin{cases}
			\sum\limits_{f\in{\cal S}}\omega_{b,f}[k]\leq \chi_{b}^{W}[k]W_{b};\\
			\sum\limits_{f\in{\cal S}}p_{b,f}[k]\leq \chi_{b}^{P}[k]P_{b}^{\max},
	\end{cases}
\end{equation}
where $W_{b}$ and $P_{b}^{\max}$ are the nominal bandwidth and maximum transmit power of O-DU $b$, respectively.
After accounting for pilot and control overhead, the number of channel uses, denoted by $n_{b,f}[k]$, assigned to flow $f$ is modeled as follows:
\begin{equation}\label{equation05}
	n_{b,f}[k]=\lfloor \eta_{b}\omega_{b,f}[k]T_{sl} \rfloor
\end{equation}
where $\eta_{b}\in(0,1)$ is the usable time-frequency fraction. For $\omega_{b,f}[k]>0$, the effective SINR, denoted by $	\gamma_{b,f}[k]$, is modeled as follows:
\begin{equation}\label{equation06}
	\gamma_{b,f}[k]=\frac{p_{b,f}[k]g_{b,f}[k]}{N_{0}\omega_{b,f}[k]+I_{b,f}[k]}
\end{equation}
where $g_{b,f}[k]$ is the effective channel power gain, $N_{0}$ is the noise power spectral density, and $I_{b,f}[k]$ is the aggregate co-channel interference. 
Assuming quasi-static channel conditions over one codeword, under the finite-blocklength normal approximation, the achievable coding rate, denoted by $R_{b,f}[k]$, in bits per channel use is modeled as:
\begin{equation}\label{equation07}
	R_{b,f}[k]\approx \left[\log_{2}\left(1+\gamma_{b,f}[k]\right)-\sqrt{\frac{V(\gamma_{b,f}[k])}{n_{b,f}[k]}}Q^{-1}(\epsilon_{b,f}[k])\right]^{+}
\end{equation}
where $\epsilon_{b,f}[k]$ denotes the target decoding error probability, $Q^{-1}(\cdot)$ is the inverse Gaussian Q-function, and $V(\gamma_{b,f}[k])$ is the channel dispersion, which is defined as follows:
\begin{equation}\label{equation08}
	V(\gamma_{b,f}[k])=\left[1-\frac{1}{(1+\gamma_{b,f}[k])^{2}}\right](\log_{2}e)^{2}.
\end{equation}

Define the scheduled information payload, denoted by $L_{b,f}[k]$, as follows:
\begin{equation}\label{equation09}
	L_{b,f}[k]=n_{b,f}[k]R_{b,f}[k]
\end{equation}
The decoding-success indicator is defined as $\xi_{b,f}[k]\in\{0,1\}$, with
\begin{equation}\label{equation10}
	\text{Pr}\left\{\xi_{b,f}[k]=1| \gamma_{b,f}[k], n_{b,f}[k] \right\} = 1-\epsilon_{b,f}[k].
\end{equation}
Therefore, we denote the successfully decoded leg-level service, denoted by $\widetilde{S}_{b,f}[k]$, as follows:
 \begin{equation}\label{equation11}
 	\widetilde{S}_{b,f}[k]=\xi_{b,f}[k]L_{b,f}[k].
 \end{equation}
The lower-layer queue, RLC, and HARQ processes map $\widetilde S_{b,s}[k]$ to packet-completion times. 

Each MCX bearer is anchored at one transmitting-side O-CU-UP/PDCP entity. 
Let $T_{s}^{\text{out}}$ denote the terminal-outcome timeout. 
Let $d_{b,s,n,u}^{\mathrm{link}}$ be the completion time of packet $n$ at receiver $u$ through O-DU $b$, as induced by the lower-layer queue/HARQ process. 
Let $\mathcal C_{s,n}(a)$ be the O-DU set used by candidate $a$. It is a singleton for SC and contains at least two O-DUs for DP.
The completion time, denoted by $d_{s,n,u}^{a}$, at receiver $u$ is given as follows:
\begin{equation}
	d_{s,n,u}^{a}=\min_{b\in\mathcal C_{s,n}(a)}
	d_{b,s,n,u}^{\mathrm{link}}
\end{equation}
Then, the group-completion time, denoted by $d_{s,n}^{a}$, is given as follows:
\begin{equation}
	d_{s,n}^{a}
	=
	\inf
	\left\{
	t:
	\frac{1}{|\mathcal U_s|}
	\sum_{u\in\mathcal U_s}
	\mathbf 1
	\left\{
	d_{s,n,u}^{a}\leq t
	\right\}
	\geq\zeta_s
	\right\},
	\label{eq:leg_group_completion}
\end{equation}
where $\zeta_s$ is the required receiver-completion fraction and $\mathcal U_s$ is the target receiver set.
The bearer-level group-completion indicator, denoted by $Y_{s,n}^{a}$, is defined as follows:
\begin{align}
	Y_{s,n}^{a}
	&=
	\mathbf 1
	\left\{
	d_{s,n}^{a}
	\leq g_{s,n}+T_s^{\mathrm{out}}
	\right\}.
	\label{eq:packet_success_compact}
\end{align} 
Thus, the unique-payload service is given by
\begin{equation}
	S_{s,a}[k]
	=
	\sum_n
	L_{s,n}
	\mathbf 1
	\left\{
	d_{s,n}^{a}\in
	[kT_{\mathrm{sl}},(k+1)T_{\mathrm{sl}})
	\right\}.
	\label{eq:unique_payload_service_compact}
\end{equation}
DP consumes resources on every selected leg but contributes each successfully delivered payload only once. Thus, a duplicated packet consumes resources on all selected links but contributes its payload to $S_{s,a}[k]$ only once. The first successfully received copy determines the packet delivery time, which is also consistent with standard packet-delay-variation measurement practice.


\subsection{Finite-Horizon Multi-QoS and Coexistence Requirements}
For each MCX flow $s\in{\cal S}_{M}$, its service requirements are summarized by:
\begin{equation}\label{equation015}
	\bm{q}_{s} \triangleq \left(\kappa_{s},\Delta_{s},r_{s}^{\min},D_{s}^{\max},J_{s}^{\max},A_{s}^{\max},T_{s}^{\text{out}},\bm{\beta}_{s},\bm{\epsilon}_{s},\zeta_{s}\right)
\end{equation}
where $\kappa_{s}$ is the service class, $r_{s}^{\min}$ is the minimum supported payload rate, $D_{s}^{\max}$, $J_{s}^{\max}$, and $A_{s}^{\max}$ are the RAN-side delay, packet-delay-variation, and AoI thresholds, respectively, $\bm{\beta}_{s}$ contains the tolerated within-window violation fractions, and $\bm{\epsilon}_{s}$ is the violation-budget vector, which are given by
\begin{equation}\label{equation016}
	\begin{cases}
			\bm{\beta}_{s}=\left[\beta_{s}^{D},\beta_{s}^{J},\beta_{s}^{A},\beta_{s}^{E}\right]^{T}; \\
				\bm{\epsilon}_{s}=\left[\epsilon_{s}^{R},\epsilon_{s}^{D},\epsilon_{s}^{J},\epsilon_{s}^{A},\epsilon_{s}^{E}\right]^{T}.
	\end{cases}
\end{equation}
where $\beta_{s}^{q}$ $(q\in{\cal Q}_{s})$ and $\epsilon_{s}^{q}$ $(q\in{\cal Q}_{s})^{+}$ are the tolerated within-window violation and the violation-budget for delay, jitter, AoI, and reliability, ${\cal Q}_{s}\subseteq\{D,J,A,E\}$ contains the QoS dimensions active for service $s$, and ${\cal Q}_{s}^{+}=\{R\}\cup {\cal Q}_{s}$ denotes the active risk-index set, where $R$ denotes the delivered-rate violation.
Setting $\beta_{s}^{q}=0$ recovers a zero-tolerance. A positive $\beta_{s}^{q}$ permits a small application-specific fraction of within-window violations.

For a successfully delivered packet, its RAN-side delay, denoted by $D_{s,n}^{a}$, is given as follows:
\begin{equation}\label{equation017}
	D_{s,n}^{a}=d_{s,n}^{a}-g_{s,n}.
\end{equation}

Let ${\cal P}_{s}^{\Delta}$ denote the set of consecutive successfully delivered packet pairs in source sequence order and $n^{-}$ be the preceding successfully delivered packet in source sequence order. For $(n^{-},n)\in {\cal P}_{s}^{\Delta}$, define the packet-delay-variation magnitude, denoted by $J_{s,n}^{a}$, as follows:
\begin{equation}\label{equation022}
J_{s,n}^{a}=\left|D_{s,n}^{a}-D_{s,n^{-}}^{a}\right|.
\end{equation}
For a status-update flow, let
\begin{equation}\label{equation023}
	u_{s}^{a}(t)=\max\left\{g_{s,n}:Y_{s,n}^{a}=1, d_{s,n}^{a}\leq t\right\}
\end{equation}
be the generation time of the freshest update delivered by time $t$. The instantaneous AoI, denoted by $AoI_{s}^{a}(t)$, is defined as follows:
\begin{equation}\label{equation024}
	AoI_{s}^{a}(t)=t-u_{s}^{a}(t).
\end{equation}
The value of $u_{s}^{a}(t_{s}^{0})$ is initialized by the freshest update delivered before the finite horizon.

Define the fully observed packet set, denoted by $\mathcal N_s^{\mathrm{out}}$,  as follows:
\begin{equation}
	\mathcal N_s^{\mathrm{out}}
	=
	\left\{
	n:t_s^0\leq 
	g_{s,n}
	\leq
	t_s^0+\Delta_s-T_s^{\mathrm{out}}
	\right\},
\end{equation}
The arrival-time definition of $\mathcal N_s^{\mathrm{out}}$ avoids outcome-dependent right censoring.
Define the successful delivered subset, denoted by $\mathcal N_{s,a}^{\mathrm{succ}}$, within the finite operational window as follows:
\begin{equation}
	\mathcal N_{s,a}^{\mathrm{succ}}
	=
	\left\{
	n\in\mathcal N_s^{\mathrm{out}}:
	Y_{s,n}^{a}=1
	\right\}.
	\label{eq:observed_packet_sets_compact}
\end{equation}
Then, the realized delivered rate and terminal non-delivery ratio are given, respectively, as follows:
\begin{align}
	\begin{cases}
	R_s^{a,\Delta}
	&=
	\frac{1}{\Delta_s}
	\sum\limits_{k\in\mathcal K_s}S_{s,a}[k];
	\\
	Z_{s,a}^{E}
	&=
	\frac{1}{|\mathcal N_s^{\mathrm{out}}|}
	\sum\limits_{n\in\mathcal N_s^{\mathrm{out}}}
	\left(1-Y_{s,n}^{a}\right),
		\end{cases}
	\label{eq:non_delivery_compact}
\end{align}
where ${\cal K}_s\triangleq\{k_{l_s},k_{l_s}+1,\ldots,k_{l_s}+K_s-1\}$ is the mission-slot index set.
The remaining mission metrics are given as follows:
\begin{align}
	\begin{cases}
	Z_{s,a}^{D} \!\!\!\!\!\!
	&=
	\frac{1}{|\mathcal N_{s,a}^{\mathrm{succ}}|}
	\sum\limits_{n\in\mathcal N_{s,a}^{\mathrm{succ}}}
	\mathbf 1\{D_{s,n}^{a}>D_s^{\max}\};
	\\
	Z_{s,a}^{J} \!\!\!\!\!\!
	&=
	\frac{1}{|\mathcal P_s^\Delta|}
	\sum\limits_{(n^-,n)\in\mathcal P_s^\Delta}
	\mathbf 1
	\left\{
	|D_{s,n}^{a}-D_{s,n^-}^{a}|>J_s^{\max}
	\right\};
	\\
	Z_{s,a}^{A} \!\!\!\!\!\!
	&=
	\frac{1}{\Delta_s}
	\int\limits_{t_s^0}^{t_s^0+\Delta_s}
	\mathbf 1
	\left\{
	AoI_{s}^{a}(t)>A_s^{\max}
	\right\}
	\,\mathrm dt .
		\end{cases}
	\label{eq:aoi_metric_compact}
\end{align}
If an active delay or jitter metric has no evaluable sample, the corresponding trajectory is conservatively counted as a mission violation.
The finite-horizon service requirements are given as follows:
\begin{equation}
	\begin{cases}
			\text{Pr}\left\{R_{s,a}^{\Delta}<r_{s}^{\min}|{\cal F}_{l_{s}}^{\text{E2}}\right\}\leq \epsilon_{s}^{R};\\
			\text{Pr}\left\{Z_{s,a}^{q}>\beta_{s}^{q}|{\cal F}_{l_{s}}^{\text{E2}}\right\}\leq \epsilon_{s}^{q}, \quad q\in {\cal Q}_{s}
	\end{cases}
\end{equation}
where $l_{s}$ is the orchestration epoch containing $t_{s}^{0}$.
Furthermore, we define the joint finite-horizon multi-QoS violation event, denoted by ${\cal V}_{s}^{MQ}$, as follows:
\begin{equation}
	{\cal V}_{s}^{MQ}=\left\{R_{s,a}^{\Delta}<r_{s}^{\min}\right\}\cup \bigcup_{q\in{\cal Q}_{s}}\left\{Z_{s,a}^{q}>\beta_{s}^{q}\right\}.
\end{equation}

For non-MCX flow $n$, define its conditional mean delivered rate, denoted by $\overline{r}_{n}[l]$, over Near-RT epoch $l$ as follows:
\begin{equation}
	\overline{r}_{n}[l]=\frac{1}{T_{x}}\mathbb{E}\left[\sum_{k\in {\cal K}_{l}}S_{n}[k]\bigg|{\cal F}_{l}^{\text{E2}}\right]
\end{equation}
where $S_{n}[k]$ is the unique payload bits for non-MCX flow $n$ in slot $k$ and ${\cal K}_{l}$ is the slot set associated with Near-RT epoch $l$.
The aggregate non-MCX utility, denoted by $U_{N}[l]$, is given as follows:
\begin{equation}\label{equation035}
	U_{N}[l]=\sum_{n\in {\cal S}_{N}} \omega_{n}\log\left(1+\frac{\overline{r}_{n}[l]}{r_{\text{ref}}}\right)
\end{equation}
where $r_{\text{ref}}>0$ is a normalization rate and $\omega_{n}>0$ is a service weight.
Let $U_{N}^{0}[l]>0$ denote the reference utility under the same traffic, channel, mobility, and infrastructure realization, but with MCX-specific reservation and prioritization disabled. The coexistence loss, denoted by ${\cal L}_{N}[l]$, is given as:
\begin{equation}
	{\cal L}_{N}[l]=\frac{\left[U_{N}^{0}[l]-U_{N}[l]\right]^{+}}{U_{N}^{0}[l]}
\end{equation}
where $(x)^{+}\triangleq\max\{0,x\}$ and the shared-network protection requirement is ${\cal L}_{N}[l]\leq \rho_{\max}$, where $\rho_{\max}\in[0,1)$.

\section{E2-Conditioned Finite-Horizon Effective Capacity}\label{sec:capability}

\subsection{Finite-Horizon Service Capability}




Let $\overline{S}_{s,a}^{\mathrm{pre}}[k]\geq 0$ denote the
backlogged potential service available to flow $s$ during the
pre-actuation phase, and let
$\overline{S}_{s,a}^{\mathrm{post}}[k]\geq 0$ denote the backlogged
potential service after candidate profile $a$ becomes effective.
For a newly admitted flow, we conservatively set
$\overline{S}_{s,a}^{\mathrm{pre}}[k]=0$ unless a default service
contract is explicitly modeled.
The cumulative actuation-aware potential service, denoted by $\overline{S}_{s,a}^{\Delta_s}$, over the complete mission horizon is defined as follows:
\begin{align}
	\overline{S}_{s,a}^{\Delta_s}
	\triangleq
	\sum_{k=k_{l_s}}^{k_{s,a}^{\mathrm{eff}}-1}
	\overline{S}_{s,a}^{\mathrm{pre}}[k]+
	\sum_{k=k_{s,a}^{\mathrm{eff}}}^{k_{l_s}+K_s-1}
	\overline{S}_{s,a}^{\mathrm{post}}[k].
	\label{eq:actuation_aware_service}
\end{align}

\textit{Definition 1 (E2-conditioned actuation-aware FH-EC):}
For QoS exponent $\theta_s>0$, the conditional service
transform, denoted by $\mathcal{M}_{s,a}^{\mathrm{FH}}(\theta_s)$, is defined as follows:
\begin{equation}\label{equation037}
	\mathcal{M}_{s,a}^{\mathrm{FH}}(\theta_s)
	\triangleq
	\mathbb{E}
	\left[
	\exp\left(
	-\theta_s\overline{S}_{s,a}^{\Delta_s}
	\right)
	\middle|
	\mathcal{F}_{l_s}^{\mathrm{E2}},\varpi_{s,a}
	\right].
\end{equation}
The corresponding E2-conditioned finite-horizon effective capacity, denoted by $C_{s,a}^{\mathrm{FH}}(\theta_s)$ is defined as:
\begin{equation}\label{equation038}
	C_{s,a}^{\mathrm{FH}}(\theta_s)
	\triangleq
	-\frac{1}{\theta_s\Delta_s}
	\log
	\mathcal{M}_{s,a}^{\mathrm{FH}}(\theta_s).
\end{equation}


Note that Definition 1 presents a finite-horizon conditional log-moment measure of the service process. Unlike conventional effective capacity, it does not replace the conditional service distribution by its long-term stationary limit~\cite{1210731,10879302}.
Instead of replacing the infinite-horizon limit by an arbitrary
observation window, the proposed FH-EC preserves the operational
horizon as a mission parameter and retains the initial conditional
distribution induced by E2 exposure.

\textit{Definition 2: (E2-conditioned finite-horizon effective bandwidth)} For $\theta_s>0$, the finite-horizon effective bandwidth, denoted by $B_{s}^{\text{FH}}(\theta_{s})$, is defined as follows:
	\begin{equation}
		B_{s}^{\text{FH}}(\theta_{s})
		\triangleq
		\frac{1}{\theta_s\Delta_s}
		\log
		\mathbb E\!\left[
		\exp\!\left(\theta_s A_s(l_s,\Delta_s)\right)
		\,\middle|\,
		\mathcal F_{l_s}^{\mathrm{E2}}		\right]
	\end{equation}
where $A_{s}(l_{s},\Delta_{s})$ is the cumulative arrival process over the same horizon, which is given by
\begin{equation}
	A_{s}(l_{s},\Delta_{s})=\sum_{k=k_{l}}^{k_{l}+K_{s}-1} A_{s}[k].
\end{equation}

Assume that, conditioned on ${\cal F}_{l}^{\text{E2}}$, the exogenous arrival process and future potential-service innovations are independent. 
For any backlog surplus threshold $q_s^{\mathrm{th}}\geq 0$, the Chernoff inequality yields:
\begin{align}\label{equation040}
&\text{Pr}\left\{A_{s}(l_{s},\Delta_{s})-\overline{S}_{s,a}^{\Delta_s}\geq q_s^{\mathrm{th}} \big | {\cal F}_{l}^{\text{E2}} ,\varpi_{s,a}\right\} \nonumber\\
&\quad 	\leq \exp\! \left[-\theta_{s}q_s^{\mathrm{th}}\!+\!\theta_{s}\Delta_{s}\left(B_{s}^{\text{FH}}(\theta_{s})\!-\!C_{s,a}^{\text{FH}}(\theta_{s})\right)\right]\!.
\end{align}
A sufficient condition for controlling the finite-window service deficit exponent is given as follows:
\begin{equation}\label{equation041}
	B_{s}^{\text{FH}}(\theta_{s})\leq C_{s,a}^{\text{FH}}(\theta_{s}).
\end{equation}
To guarantee a prescribed service-deficit probability, denoted by $\epsilon_{s}^{sd}\in(0,1)$, a sufficient feasibility condition is given as follows:
\begin{equation}\label{equation042}
	B_{s}^{\text{FH}}(\theta_{s})+\frac{\log(1/\epsilon_{s}^{sd})}{\theta_{s}\Delta_{s}}\leq C_{s,a}^{\text{FH}}(\theta_{s})+\frac{q_s^{\mathrm{th}}}{\Delta_{s}}.
\end{equation}
Thus, Eq.~\eqref{equation041} preserves the traditional EC interpretation, whereas Eq.~\eqref{equation042} supplies an explicit finite-window probability margin. 




To obtain a computable representation of Eq.~\eqref{equation037}, let
\begin{equation}
	X_{s,a}[k]
	=
	\left(
	\{Z_{s,a,\ell}[k]\}_{\ell\in\mathcal L_a},
	Z_{s,a}^{\mathrm{cm}}[k],
	H_{s,a}[k]
	\right)
	\in\mathcal X_{s,a},
\end{equation}
denote a finite-state abstraction of the domain condition, where $Z_{s,a,\ell}[k]$ and $Z_{s,a}^{\mathrm{cm}}[k]$ are the radio/service state over leg $l$ and transport, O-CU-UP, shared power, or common-mode failure state, respectively, $\mathcal L_a$ is the duplication-leg index set, and $H_{s,a}[k]$ is the HARQ/retransmission phase.
The state can jointly represent quantized channel quality, interference/load level, O-DU availability, retransmission phase, and other variables governing the service process. Finite-state Markov channel models are a standard abstraction for temporally correlated wireless channels~\cite{350282}.

The current conditional state distribution, denoted by $\bm{\alpha}_{s,a}[l]$, is given as follows:
\begin{equation}
	\bm{\alpha}_{s,a}[l]=\left[\text{Pr}\left\{X_{s,a}[k_{l}]=m \big | {\cal F}_{l}^{\text{E2}} \right\}\right]_{m=1}^{M_{s,a}}.
\end{equation}
For phase $r\in\{\mathrm{pre},\mathrm{post}\}$, we define the
phase-dependent service-weighted transition kernel, denoted by $\left[\mathbf{G}_{s,a}^{r}[k](\theta_s)
\right]_{uv}$, as follows:
\begin{align}
	&\left[\mathbf{G}_{s,a}^{r}[k](\theta_s)
	\right]_{uv}\triangleq
	\mathbb{E}
	\Bigg[
	\exp\left(
	-\theta_s\overline{S}_{s,a}^{r}[k]
	\right)
	\nonumber\\
	&\quad\times 
	\mathbf{1}\{X_{s,a}[k+1]=v\}
	\bigg|
	X_{s,a}[k]=u,
	\mathcal{F}_{l_s}^{\mathrm{E2}}
	\varpi_{s,a}^{r}
	\Bigg],
	\label{eq:phase_kernel}
\end{align}
where
$\varpi_{s,a}^{\mathrm{post}}\equiv\varpi_{s,a}$ and
$\varpi_{s,a}^{\mathrm{pre}}\equiv\varpi_s^{\mathrm{pre}}$.
For compactness, we define the following time-ordered product:
\begin{equation}
	\prod\limits_{k=a}^{\longrightarrow b}\bm{G}[k]\triangleq \bm{G}[a]\bm{G}[a+1]\cdots \bm{G}[b].
\end{equation}
Define the pre-actuation kernel product, denoted by $\mathbf{H}_{s,a}^{\mathrm{pre}}(\theta_s)$, and post-actuation kernel product, denoted by $\mathbf{H}_{s,a}^{\mathrm{post}}(\theta_s)$, respectively, as follows:
\begin{equation}
	\begin{cases}
	\mathbf{H}_{s,a}^{\mathrm{pre}}(\theta_s)\!\!\!\!
	&\triangleq\!\!\!\!\!\!\!\!\!\!\!\!
	\prod\limits_{\quad k=k_{l_s}}^{\qquad\,\,\,\,\rightarrow k_{s,a}^{\mathrm{eff}}-1}
	\mathbf{G}_{s,a}^{\mathrm{pre}}[k](\theta_s);\\
	\mathbf{H}_{s,a}^{\mathrm{post}}(\theta_s)\!\!\!\!
	&\triangleq\!\!\!\!\!\!\!\!\!\!\!\!\!\!\!\!\!\!\!
		\prod\limits_{\quad k=k_{s,a}^{\mathrm{eff}}}^{\qquad\quad\,\,\,\,\rightarrow 
			k_{l_s}+K_s-1}\!\!\!\!\!\!\!
	\mathbf{G}_{s,a}^{\mathrm{post}}[k](\theta_s).
		\end{cases}
	\label{eq:pre_kernel_product}
\end{equation}
An empty product is defined as the identity matrix.
Then, the exact actuation-aware FH-EC for the adopted finite-state
Markov-additive model is therefore given as:
\begin{align}
	C_{s,a}^{\mathrm{FH}}(\theta_s)
	=
	-\frac{1}{\theta_s\Delta_s}
	\log
	\Big[
	\boldsymbol{\alpha}_{s,a}^{T}[l_s]\,
	\mathbf{H}_{s,a}^{\mathrm{pre}}(\theta_s)\,
	\mathbf{H}_{s,a}^{\mathrm{post}}(\theta_s)\,
	\mathbf{1}
	\Big].
	\label{eq:matrix_actuation_fhec}
\end{align}


\textit{Proposition 1 (Long-horizon consistency)}
Fix profile $a$ and exponent $\theta_s>0$. For a sequence of
mission horizons $\Delta_s=K_sT_{\rm sl}$, let
$d_{K_s}$ denote the number of pre-actuation slots and
$\mathbf H_{K_s}^{\rm pre}$ be the corresponding pre-actuation
kernel product. Suppose that:

1) $\sup_{K_s} d_{K_s}\le d_{\max}<\infty$;

2) After actuation, the service-weighted kernel is
time homogeneous and primitive:
\begin{equation}
\mathbf G_{s,a}^{\rm post}[k](\theta_s)
=
\mathbf G_{s,a}^{\infty}(\theta_s),
\qquad k\ge k_{s,a}^{\rm eff};
\end{equation}

3) If $\bm{v}_{s,a}$ is the positive Perron right
eigenvector of $\mathbf G_{s,a}^{\infty}(\theta_s)$, then
there exist constants $0<c_-\le c_+<\infty$ such that
\begin{equation}
c_-
\le
\boldsymbol\alpha_{s,a}^{T}[l_s]
\mathbf H_{K_s}^{\rm pre}(\theta_s)
\bm{v}_{s,a}
\le
c_+,
\qquad \forall K_s.
\end{equation}
Then, the long-horizon effective capacity, denoted by $C_{s,a}^{\infty}(\theta_s)$, is obtained as follows:
\begin{equation}
	C_{s,a}^{\infty}(\theta_s)\triangleq \lim_{K_s\rightarrow\infty}
	C_{s,a}^{\rm FH}(\theta_s)
	=
	-\frac{1}{\theta_sT_{\rm sl}}
	\log\rho\!\left(
	\mathbf G_{s,a}^{\infty}(\theta_s)
	\right).
\end{equation}
where $\rho(\cdot)$ denotes the spectral radius.

\begin{IEEEproof}
	The proof of Proposition~1 is in Appendix A.
\end{IEEEproof}

\noindent\textbf{Remark 1:} Proposition 1 shows that the finite E2 selection-and-actuation transient affects short mission windows but contributes only a vanishing $O(d_{s,a}^{\mathrm{ctl}}/K_s)$ term in the long-horizon regime. Consequently, the proposed FH-EC retains the conventional effective-capacity limit without ignoring the service loss before an E2 action becomes effective.

Assume that the admitted MCX traffic satisfies the following finite-window arrival envelope:
\begin{equation}
	A_s(l_s,\Delta_s)
	\leq
	\sigma_s+r_s\Delta_s,
	\label{eq:arrival_envelope}
\end{equation}
where $\sigma_s$ is the maximum admitted burst and $r_s$ is the
sustained payload rate.
For any service-deficit threshold $q_s^{\mathrm{th}}\geq0$, by applying Markov's inequality, we have
\begin{align}
	&
	\Pr
	\left\{
	A_s(l_s,\Delta_s)-\overline{S}_{s,a}^{\Delta_s}
	\geq q_s^{\mathrm{th}}
	\mid
	\mathcal F_{l_s}^{E2},\varpi_{s,a}
	\right\}
	\notag\\
	&\quad \leq
	\exp
	\left[
	-\theta_s(q_s^{\mathrm{th}}-\sigma_s)
	+
	\theta_s\Delta_s
	\left(
	r_s-C_{s,a}^{\mathrm{FH}}(\theta_s)
	\right)
	\right].
	\label{eq:fh_service_deficit_bound}
\end{align}

\textit{Definition 3: (Finite-horizon supportable admitted rate)} For a prescribed service-deficit probability $\epsilon_{s}^{sd}$, candidate $a$ supports any rate not larger than the finite-horizon supportable admitted rate, denoted by $r_{s,a}^{\mathrm{FH}}$, which is given as follows:
	\begin{equation}
		r_{s,a}^{\mathrm{FH}}
		=\left[
		C_{s,a}^{\mathrm{FH}}(\theta_s)
		+
		\frac{q_s^{\mathrm{th}}-\sigma_s}{\Delta_s}
		-
		\frac{\log(1/\epsilon_{s}^{sd})}
		{\theta_s\Delta_s}\right]^{+}.
		\label{eq:fh_supportable_rate}
	\end{equation}

\subsection{Capability Characterization under Connectivity Diversity}
The objective is not to design a new duplication mechanism,
but to quantify heterogeneous connectivity options through
the proposed FH-EC primitive. 
For a fixed E2-executable policy $\varpi_{s,a}$, we define the delivered-rate risk and the dimension-specific multi-QoS risk, respectively, as follows:
\begin{equation}\label{equation065}
	\begin{cases}
p_{s,a}^{R,\text{FH}}\triangleq \text{Pr}\left\{R_{s,a}^{\Delta}<r_{s,a}\big| {\cal F}_{l}^{\text{E2}},\varpi_{s,a}\right\};\\
	p_{s,a}^{q,\text{FH}}\triangleq \text{Pr}\left\{Z_{s,a}^{q}>\beta_{s}^{q}\big| {\cal F}_{l}^{\text{E2}},\varpi_{s,a}\right\}.
	\end{cases}
\end{equation}
where $r_{s,a}\geq r_{s}^{\min}$ is the finite-horizon delivered-rate target.
The associated multi-QoS risk vector, denoted by $\bm{p}_{s,a}^{\text{FH}}$, is given by 
\begin{equation}
	\bm{p}_{s,a}^{\text{FH}}=\left[p_{s,a}^{R,\text{FH}},\left\{p_{s,a}^{q,\text{FH}}\right\}_{q\in{\cal Q}_{s}}\right]^{T}.
\end{equation}
An exact Markov recursion could in principle be constructed by augmenting the state with accumulated violation counters, the number of evaluable packets or packet pairs, and the accumulated AoI violation time.
However, this would produce a high-dimensional state whose size grows with the mission horizon and packet count. We therefore estimate Eq.~\eqref{equation065} from complete finite-horizon delivery trajectories rather than claiming a low-dimensional exact recursion.

\subsubsection{Lower Confidence Bound on Finite-Horizon Capability}
Let $c_l=g\!\left(\mathcal F_l^{\rm E2}\right)\in\mathcal C$ denote the finite E2-context label, where the context map $g$, the finite context set $\mathcal C$, and the candidate-policy library are fixed using an independent calibration dataset $\mathcal D^{\rm tr}$ before $\mathcal D^{\rm cal}$ is accessed.
For a concurrent profile-selection vector $\bar{\bm{x}}$,
let $e_{s,a}(\bar{\bm{x}},c)\in\mathcal E_c^{\rm exec}$ denote the finite execution-class label experienced by profile $a$ of flow $s$ in context cell $c$. The execution class records the active MCX service composition, aggregate resource contract, participating radio legs, fixed O-CU/O-DU policies, calibrated interference/load envelope, and common-mode failure class. The finite execution-class set $\mathcal E_c^{\rm exec}$ is fixed before calibration.
 Define the calibration key as $\jmath\triangleq\left(	a,c,e,\theta_s,\bm{q}_s\right)\in\mathcal J$, where $\bm{q}_s$ contains the service class, mission horizon,
and all active delivered-rate, delay, jitter, AoI, and terminal
non-delivery requirements. The finite calibration library
$\mathcal J$ is fixed before $\mathcal D^{\rm cal}$ is accessed.

Let $\mathcal W_{\jmath}$ denote a complete mission trajectory
generated under calibration key $\jmath$. Then, the exact
cell-and-execution-class-conditioned FH-EC is given by
\begin{equation}
	C_{\jmath}^{\rm FH}
	\triangleq
	-\frac{1}{\theta_s\Delta_s}
	\log
	\mathbb E
	\left[
	\exp\left(
	-\theta_s\overline S_{\jmath}^{\Delta_s}
	\right)
	\mid
	c_{l_s}=c,\ e_{s,a}=e
	\right].
\end{equation}
For $q\in\mathcal Q_s^{+}$, define
\begin{equation}
	p_{\jmath}^{q,\rm FH}
	\triangleq
	\Pr
	\left\{
	\mathcal V_{\jmath}^{q}
	\mid
	c_{l_s}=c,\ e_{s,a}=e
	\right\},
\end{equation}
where
\begin{equation}
	\mathcal V_{\jmath}^{R}
	=
	\left\{
	R_{s,a}^{\Delta_s}<r_s^{\rm tgt}
	\right\},
	\qquad
	\mathcal V_{\jmath}^{q}
	=
	\left\{
	Z_{s,a}^{q}>\beta_s^q
	\right\},
	\quad q\in\mathcal Q_s .
\end{equation}

Suppose that $N_\jmath$ independent complete mission trajectories are available for profile $\jmath$, producing accumulated potential services $\overline S_\jmath^{(1)}(\Delta_s), \ldots, \overline S_\jmath^{(N_\jmath)}(\Delta_s)$.
We define 
\begin{equation}
	Y_\jmath^{(n)} = \exp \left( -\theta_s \overline S_\jmath^{(n)}(\Delta_s) \right) \in(0,1]
\end{equation}
and 
\begin{equation}
	\widehat\mu_\jmath = \frac{1}{N_\jmath} \sum_{n=1}^{N_\jmath} Y_\jmath^{(n)}.
\end{equation}
For a confidence budget $\alpha_\jmath^C\in(0,1)$, by applying  Hoeffding's inequality, we have~\cite{Hoeffding01031963}
\begin{equation}
	\Pr_{\mathrm{cal}} \left\{ \mathbb E[Y_\jmath] \leq \widehat\mu_\jmath + \sqrt{ \frac{ \log(1/\alpha_\jmath^C) }{ 2N_\jmath } } \right\} \geq 1-\alpha_\jmath^C.
\end{equation}
Because $-\log(\cdot)$ is decreasing, a valid lower confidence bound on the finite-horizon capability is
\begin{equation}
 \underline C_\jmath^{\mathrm{FH}} = -\frac{1}{\theta_s\Delta_s} \log \left[ \min \left\{ 1,\, \widehat\mu_\jmath + \sqrt{ \frac{ \log(1/\alpha_\jmath^C) }{ 2N_\jmath } } \right\} \right]. 
\end{equation}
Hence, we have
\begin{equation}
	\Pr_{\mathrm{cal}} \left\{ C_\jmath^{\mathrm{FH}} \geq \underline C_\jmath^{\mathrm{FH}} \right\} \geq 1-\alpha_\jmath^C.
\end{equation}

\subsubsection{Statistically Calibrated Executable Profiles}
The exact finite-horizon capability and operational risks in
$\boldsymbol{\phi}_{s,a}[l]$ are generally unknown and must be inferred from finite data. To separate policy construction from statistical validation, the context partition, candidate-policy library, and all model hyperparameters are fixed using a training set, denoted by $\mathcal{D}^{\mathrm{tr}}$ before the independent calibration set $\mathcal{D}^{\mathrm{cal}}$ is accessed.

Define the simultaneous profile-validity event, denoted by $\mathcal E_{\rm prof}$, as follows:
\begin{equation}
	\mathcal E_{\rm prof}
	\triangleq
	\bigcap_{\jmath\in\mathcal J}
	\left\{
	C_{\jmath}^{\rm FH}
	\ge
	\underline C_{\jmath}^{\rm FH},
	\quad
	p_{\jmath}^{q,\rm FH}
	\le
	\overline p_{\jmath}^{q,\rm FH},
	\ \forall q\in\mathcal Q_{\jmath}^{+}
	\right\}.
	\label{eq:Eprof-new}
\end{equation}
where $\overline{p}_\jmath^q$ is the exact Clopper–Pearson upper confidence limit, which is given as follows:
\begin{equation}
	\overline{p}_\jmath^q = \begin{cases} 1, & K_{\jmath,q}=N_\jmath; \\[1mm] \operatorname{Beta}^{-1}\! \left( 1\!-\!\alpha_{\jmath,q}; K_{\jmath,q}\!+\!1, N_\jmath-K_{\jmath,q} \right), & K_{\jmath,q}<N_\jmath, \end{cases}
\end{equation}
where $\operatorname{Beta}^{-1}(\cdot)$ is the beta distribution, $\alpha_{\jmath,q}$ is the one-sided confidence budget, and $K_{\jmath,q}$ is the observed number of type-$q$ mission-window violations.

Assigning the one-sided confidence-failure budgets, we have
\begin{equation}\label{equation066a}
	\sum_{\jmath\in\mathcal J}
	\left(
	\alpha_{\jmath}^{C}
	+
	\sum_{q\in\mathcal Q_{\jmath}^{+}}
	\alpha_{\jmath,q}
	\right)
	\le
	\alpha_{\rm prof}.
\end{equation}
By applying the individual one-sided coverage guarantees and the union
bound, we obtain
\begin{equation}
	\Pr_{\rm cal}
	\left\{
	\mathcal E_{\rm prof}
	\right\}
	\ge
	1-\alpha_{\rm prof}.
\end{equation}
Because $\mathcal E_{\rm prof}$ holds simultaneously over the
entire finite library $\mathcal J$, subsequently selecting a
calibrated profile after observing the current E2-context cell
does not incur an additional post-selection confidence penalty.
This guarantee is conditional on the calibrated context and
execution class; it is not a pointwise guarantee for every raw
realization of $\mathcal F_l^{\rm E2}$ within the same context cell.

\subsubsection{Confidence-Calibrated E2-Executable Profiles}

For calibration key $\jmath$, we define the confidence-calibrated executable profile, denoted by $\phi_{\jmath}$, as follows:
\begin{equation}
	\phi_{\jmath}
	\triangleq
	\left(
	\jmath,
	m_a,
	\mathcal I_a,
	\bm{b}_{s,a},
	\varpi_{s,a},
	\underline C_{\jmath}^{\rm FH},
	\overline{\bm{p}}_{\jmath}^{\rm FH},
	\tau_{s,a}^{\rm ctl}
	\right),
	\label{eq:calibrated-profile-new}
\end{equation}
where
\begin{equation}
	\overline{\bm{p}}_{\jmath}^{\rm FH}
	=
	\left[
	\overline p_{\jmath}^{R,\rm FH},
	\left\{
	\overline p_{\jmath}^{q,\rm FH}
	\right\}_{q\in\mathcal Q_s}
	\right]^T .
\end{equation}


Let ${\cal A}_{s}^{\text{E2}}[l]$ denote the executable candidate set.
At the admission epoch $l_s$, let $c_{l_s}$ denote the calibrated E2-context cell and let ${\cal C}_{s,a}$ denote the set of context cells for which profile $a$ has been calibrated. 
The capability-feasible profile set, denoted by $\bm{ \Phi}_{s}^{\text{FH}}[l]$, representing the finite candidate set, is given as follows:
\begin{align}
	\Phi_s^{\mathrm{FH}}[l_s]
	\triangleq
	\Big\{
	\phi_{\jmath}:\
	&a\in\mathcal{A}_s^{\mathrm{E2}}[l_s],
	\quad c_{l_s}\in{\cal C}_{s,a},
	\nonumber\\
	&d_{s,a}^{\mathrm{ctl}}<K_s,
	\quad r_{s,a}^{\mathrm{FH}}\geq r_s^{\min},
	\nonumber\\
	&B_s^{\mathrm{FH}}(\theta_s)
	\leq
	\underline{C}_{s,a}^{\mathrm{FH}}(\theta_s),
	\nonumber\\
	&B_s^{\mathrm{FH}}(\theta_s)
	+
	\frac{\log(1/\epsilon_s^{\mathrm{sd}})}
	{\theta_s\Delta_s}
	\leq
	\underline{C}_{s,a}^{\mathrm{FH}}(\theta_s)
	+
	\frac{q_s^{\mathrm{th}}}{\Delta_s},
	\nonumber\\
	&\overline{p}_{s,a}^{q,\mathrm{FH}}\leq\epsilon_s^q,
	\quad q\in\mathcal{Q}_{s}^{+},
	\nonumber\\
	&\tau_{s,a}^{\mathrm{ctl}}
	\leq
	\min\{T_x,\Delta_s\}
	\Big\}.
	\label{eq:revised_feasible_profiles}
\end{align}

\section{FH-EC-Driven MCX Profile Orchestration in O-RAN}

\subsection{Joint Profile Selection Under Shared Resources}

At orchestration epoch $l$, partition the active MCX flows into the following sets:
\begin{equation}
	\begin{cases}
	\mathcal{S}_M[l]
	=
	\mathcal{S}_M^{\mathrm{lock}}[l]
	\cup
	\mathcal{S}_M^{\mathrm{new}}[l];\\
	\mathcal{S}_M^{\mathrm{lock}}[l]
	\cap
	\mathcal{S}_M^{\mathrm{new}}[l]
	=
	\emptyset,
		\end{cases}
	\label{eq:flow_partition}
\end{equation}
where $\mathcal{S}_M^{\mathrm{lock}}[l]$ contains previously
admitted flows whose mission windows have not expired, and
$\mathcal{S}_M^{\mathrm{new}}[l]$ contains flows requesting
admission at epoch $l$.
For every $s\in\mathcal{S}_M^{\mathrm{lock}}[l]$, let
$a_s^{\mathrm{lock}}$ denote its profile selected at admission.
The corresponding profile-selection variable is not re-optimized
during $\mathcal{T}_s$. The aggregate resource contract, denoted by $\boldsymbol{b}^{\mathrm{lock}}[l]$, of the locked flows is given as follows:
\begin{equation}
	\boldsymbol{b}^{\mathrm{lock}}[l]
	\triangleq
	\sum_{s\in\mathcal{S}_M^{\mathrm{lock}}[l]}
	\boldsymbol{b}_{s,a_s^{\mathrm{lock}}}.
	\label{eq:locked_resources}
\end{equation}
For each new flow $s\in\mathcal{S}_M^{\mathrm{new}}[l]$, define
the eligible nonzero candidate set, denoted by $\mathcal{A}_s^{+}[l]$, as follows:
\begin{equation}
	\mathcal{A}_s^{+}[l]
	\triangleq
	\left\{
	a:\phi_{\jmath}\in\Phi_s^{\mathrm{FH}}[l]
	\right\}.
	\label{eq:new_candidate_set}
\end{equation}
A zero profile $a=0$ is added only for a new flow, with
$\boldsymbol{b}_{s,0}=\boldsymbol{0}$ and
$V_{s,0}=0$. Hence, we have $\mathcal{A}_s[l]=\mathcal{A}_s^{+}[l]\cup\{0\}$, $	s\in\mathcal{S}_M^{\mathrm{new}}[l]$.
The zero profile denotes admission rejection and cannot be used to
withdraw a previously certified mission-horizon contract.


Let $c_l$ denote the calibrated E2-context cell at epoch $l$.
The context descriptor includes the active MCX service composition used to construct the profile library.  Let
\begin{equation}
	\bm{B}^{\text{av}}[l]=\left[B_{1}^{\text{av}}[l],\dots,B_{R}^{\text{av}}[l]\right]^{T}\in \mathbb{R}_{+}^{R}
\end{equation}
denote the resource vector available at epoch $l$, after accounting for O-DU degradation and resources reserved outside the considered orchestration problem.
Define $\mathcal G_c$ as the finite set of aggregate MCX resource contracts attainable in context cell $c$.
We have $\mathbf 0\in\mathcal G_c$. 
Fix a non-MCX allocation rule
$\psi_{\mathrm N}$ that is used during both calibration and runtime
execution. For an aggregate MCX contract
$\bm{b}\in\mathcal G_c$, let $	\overline r_{n,c}^{\psi_{\mathrm N}}(\bm{b})$
denote the true conditional mean rate of non-MCX flow $n$ when
$\bm{b}$ is enforced in context cell $c$ and
$\psi_{\mathrm N}$ allocates the residual resources.
We define the corresponding non-MCX utility, denoted by $U_{\mathrm N,c}^{\psi_{\mathrm N}}(\bm{b})$, as follows:
\begin{equation}
	U_{\mathrm N,c}^{\psi_{\mathrm N}}(\bm{b})
	\triangleq
	\sum_{n\in\mathcal S_{\mathrm N}}
	\omega_n
	\log
	\left(
	1+
	\frac{
		\overline r_{n,c}^{\psi_{\mathrm N}}(\bm{b})
	}{
		r_{\mathrm{ref}}
	}
	\right).
	\label{eq:true-nonmcx-context-utility}
\end{equation}
Let $U_{\mathrm N,c}^{0}$ denote the context-matched reference utility with MCX-specific reservation and prioritization disabled.
The policy-induced coexistence-supportable set, denoted by $\mathcal B_{\mathrm N,c}^{\psi_{\mathrm N}}$, is given by
\begin{align}
	\mathcal B_{\mathrm N,c}^{\psi_{\mathrm N}}
	\triangleq
	\Bigg\{
	\bm{b}\in\mathcal G_c:\quad&
	U_{\mathrm N,c}^{\psi_{\mathrm N}}(\bm{b})
	\geq
	(1-\rho_{\max})U_{\mathrm N,c}^{0},
	\nonumber\\[-1mm]
	&
	\overline r_{n,c}^{\psi_{\mathrm N}}(\bm{b})
	\geq
	\underline r_n^{\mathrm N},
	\quad
	n\in\mathcal S_{\mathrm N}^{\mathrm{prot}}
	\Bigg\}.
	\label{eq:policy-induced-coexistence-set}
\end{align}
Since $\psi_{\mathrm N}$ provides an explicit feasible non-MCX
allocation, we have
\begin{equation}
	\mathcal B_{\mathrm N,c}^{\psi_{\mathrm N}}
	\subseteq
	\mathcal B_{\mathrm N,c}^{\mathrm{coex}}
	\cap
	\mathcal G_c
	\label{eq:policy-set-inside-exact-set}
\end{equation}
where $B_{\mathrm N,c}^{\mathrm{coex}}$ is the exact set that satisfies the non-MCX requirements given the existence of a feasible residual policy.
For each context cell $c$ and aggregate MCX contract
$\bm{b}\in\mathcal G_c$, let $\left\{R_{n,c}^{(j)}(\bm{b})\right\}_{j=1}^{N_{c,\bm{b}}}$ be independent calibration realizations of the epoch-level delivered
rate of non-MCX flow $n$ under the fixed residual-resource policy
$\psi_N$. Assume that
\begin{equation}
	0\leq R_{n,c}^{(j)}(\bm{b})\leq r_n^{\max},
	\label{eq:nonmcx_rate_bound}
\end{equation}
where $r_n^{\max}$ is a known physical upper bound.
The sample mean, denoted by $\widehat r_{n,c}(\bm{b})$, is derived as follows:
\begin{equation}
	\widehat r_{n,c}(\bm{b})
	=
	\frac{1}{N_{c,\bm{b}}}
	\sum_{j=1}^{N_{c,\bm{b}}}
	R_{n,c}^{(j)}(\bm{b}).
	\label{eq:nonmcx_rate_sample_mean}
\end{equation}
For a one-sided confidence budget
$\alpha_{n,c,\bm{b}}^{-}$, Hoeffding's inequality gives the following lower confidence bound (LCB):
\begin{equation}
	\underline r_{n,c}(\bm{b})
	=
	\left[
	\widehat r_{n,c}(\bm{b})
	-
	r_n^{\max}
	\sqrt{
		\frac{\log(1/\alpha_{n,c,\bm{b}}^{-})}
		{2N_{c,\bm{b}}}
	}
	\right]^+ .
	\label{eq:nonmcx_rate_lcb}
\end{equation}
Likewise, let
$\{R_{n,c}^{0,(j)}\}_{j=1}^{N_c^0}$ denote the context-matched
reference-rate realizations with MCX-specific reservation and
prioritization disabled. We define
\begin{equation}
	\widehat r_{n,c}^{0}
	=
	\frac{1}{N_c^0}
	\sum_{j=1}^{N_c^0}
	R_{n,c}^{0,(j)}
	\label{eq:baseline_rate_mean}
\end{equation}
and the upper confidence bound (UCB) as follows:
\begin{equation}
	\overline r_{n,c}^{0}
	=
	\min\left\{
	r_n^{\max},
	\widehat r_{n,c}^{0}
	+
	r_n^{\max}
	\sqrt{
		\frac{\log(1/\alpha_{n,c}^{0})}
		{2N_c^0}
	}
	\right\}.
	\label{eq:baseline_rate_ucb}
\end{equation}
Since the utility function is coordinate-wise increasing, define
\begin{align}
	\begin{cases}
	\underline U_{N,c}(\bm{b})
	&=
	\sum_{n\in\mathcal S_N}
	\omega_n
	\log\left(
	1+\frac{\underline r_{n,c}(\bm{b})}{r^{\mathrm{ref}}}
	\right);
	\\
	\overline U_{N,c}^{0}
	&=
	\sum_{n\in\mathcal S_N}
	\omega_n
	\log\left(
	1+\frac{\overline r_{n,c}^{0}}{r^{\mathrm{ref}}}
	\right).
		\end{cases}
	\label{eq:baseline_utility_ucb}
\end{align}
Assign the confidence budgets such that
\begin{equation}\label{equation087}
	\sum_c
	\left[
	\sum_{n\in\mathcal S_N}\alpha_{n,c}^{0}
	+
	\sum_{\bm{b}\in\mathcal G_c}
	\sum_{n\in\mathcal S_N}
	\alpha_{n,c,\bm{b}}^{-}
	\right]
	\leq \alpha_N.
\end{equation}
We define the simultaneous non-MCX validity event, denoted by $\mathcal E_N$, as follows:
\begin{equation}
	\begin{aligned}
		\mathcal E_N
		\triangleq
		\bigcap_c
		\Bigg[
		&
		\bigcap_{\bm{b}\in\mathcal G_c}
		\bigcap_{n\in\mathcal S_N}
		\left\{
		r_{n,c}^{\psi_N}(\bm{b})
		\geq
		\underline r_{n,c}(\bm{b})
		\right\}
		\\
		&\cap
		\bigcap_{n\in\mathcal S_N}
		\left\{
		r_{n,c}^{0}
		\leq
		\overline r_{n,c}^{0}
		\right\}
		\Bigg].
	\end{aligned}
	\label{eq:nonmcx_validity_event}
\end{equation}
By Hoeffding's inequality and the union bound, we obtain
\begin{equation}
	\Pr_{\mathrm{cal}}\{\mathcal E_N\}
	\geq 1-\alpha_N.
	\label{eq:nonmcx_validity_probability}
\end{equation}
Thus, the finite-sample calibration-safe set, denoted by $\widehat{\mathcal G}_{c}^{\mathrm{safe}}$, is derived as follows:
\begin{equation}
	\begin{aligned}
		\widehat{\mathcal G}_{c}^{\mathrm{safe}}
		\triangleq
		\Big\{
		\bm{b}\in\mathcal G_c:
		&
		\ \underline U_{N,c}(\bm{b})
		\geq
		(1-\rho^{\max})\overline U_{N,c}^{0},
		\\
		&
		\ \underline r_{n,c}(\bm{b})
		\geq r_n^N,
		\quad
		n\in\mathcal S_N^{\mathrm{prot}}
		\Big\}.
	\end{aligned}
	\label{eq:finite_sample_safe_set}
\end{equation}
The origin-anchored calibration-safe MCX budget, denoted by $\bm{b}_{c}^M$, is selected as follows:
\begin{equation}
	\begin{aligned}
		\bm{b}_{c}^M
		\in
		\underset{\bm{b}\in\mathcal G_c}{\arg\max}
		\quad
		& \boldsymbol w^{\mathrm T}\bm{b}
		\\
		\mathrm{s.t.}\quad
		&
		\left\{
		\bm{b}\in\mathcal G_c:
		\mathbf 0\preceq\bm{b}\preceq\bm{b}
		\right\}
		\subseteq
		\widehat{\mathcal G}_{c}^{\mathrm{safe}}
	\end{aligned}
	\label{eq:origin_anchored_safe_budget_revised}
\end{equation}
where $\boldsymbol w \succeq 0$ is the resource-box weight.
If no nonzero budget satisfies Eq.~\eqref{eq:origin_anchored_safe_budget_revised},
the zero budget is used.
On $\mathcal E_N$, we have
\begin{equation}
	\left\{
	\bm{b}\in\mathcal G_c:
	\mathbf 0\preceq\bm{b}\preceq\bm{b}_{c}^M
	\right\}
	\subseteq
	\mathcal B_{N,c}^{\psi_N}
	\subseteq
	\mathcal B_{N,c}^{\mathrm{coex}}.
	\label{eq:safe_budget_inclusion_revised}
\end{equation}
Hence, the runtime constraint $\bm{b}^{\mathrm M}(\bm{x})\preceq\overline{\bm{b}}_{c_l}^{\mathrm M}$ is a calibration-supported inner restriction of the exact coexistence region over the finite set of attainable profile contracts.

\subsection{Joint Profile-Selection Problem and Finite-Library Profile Selection}

For every confidence-screened nonzero profile, we define the
conservative FH-EC surplus as follows:
\begin{equation}
	\Delta\underline{C}_{\jmath}^{\mathrm{FH}}
	\triangleq
	\underline{C}_{\jmath}^{\mathrm{FH}}(\theta_s)
	-
	B_s^{\mathrm{FH}}(\theta_s)
	\geq 0.
	\label{eq:conservative_surplus}
\end{equation}
The secondary profile value, denoted by $V_{s,\jmath}$, is given by
\begin{equation}
	V_{s,\jmath}
	=
	\omega_s
	\Delta\underline{C}_{\jmath}^{\mathrm{FH}}
	-
	\lambda_b
	\left\|
	\widetilde{\boldsymbol{b}}_{s,a}
	\right\|_1
	-
	\lambda_\tau
	\tau_{s,a}^{\mathrm{ctl}},
	\label{eq:revised_profile_value}
\end{equation}
where $\widetilde{\boldsymbol{b}}_{s,a}$ is the component-wise
normalized resource vector. Thus, $\lambda_b$ penalizes resource
consumption rather than profile reconfiguration.
For every new flow $s\in\mathcal{S}_M^{\mathrm{new}}[l]$ and
$a\in\mathcal{A}_s[l]$, let
$x_{s,\jmath}[l]\in\{0,1\}$ indicate profile selection. 
For every candidate profile, define
\begin{align}
	\overline x_{s,\jmath}[l]
	\triangleq
	\begin{cases}
		1,
		& s\in\mathcal S_M^{\rm lock}[l],\\
		x_{s,\jmath}[l],
		& s\in\mathcal S_M^{\rm new}[l],\\
		0, & \text{otherwise}.
	\end{cases}
\end{align}
Define the total MCX contract after admitting the new flows as:
\begin{align}
	\bm{b}^M(\bm{x};l)
	=
	\sum_{s}
	\sum_{\jmath\in\mathcal J}
	\bm{b}_{\jmath}\,
	\overline x_{s,\jmath}[l].
\end{align}
To prioritize MCX admission over secondary resource efficiency, we
define
\begin{equation}
	\widetilde{V}_{s,\jmath}
	\triangleq
	\Gamma\,\omega_s^{\mathrm{adm}}\mathbf{1}\{a\neq0\}
	+
	V_{s,\jmath},
	\label{eq:admission_priority_value}
\end{equation}
where $\Gamma>0$ is selected sufficiently large so that admission
priority dominates the secondary profile value and $\omega_s^{\mathrm{adm}}$ is the admission priority. Equivalently,
Eq.~\eqref{eq:admission_priority_value} may be implemented as a
lexicographic two-stage objective.
Let $\mathcal H_l$ denote the precomputed set of minimal
forbidden profile combinations at epoch $l$. A profile
combination belongs to $\mathcal H_l$ if its induced execution
class is absent from the calibrated library $\mathcal J$, or if
it violates a stored resource, interference, load, policy, or
common-mode-failure envelope. 

The FH-EC profile orchestration problem is formulated as follows:
\begin{subequations}
	\label{prob:revised_p1}
	\begin{align}
		\mathbf{P1}[l]:
		\max_{\{x_{s,\jmath}[l]\}}&
		\sum_{s\in\mathcal{S}_M^{\mathrm{new}}[l]}
		\sum_{\jmath\in\mathcal{J}}
		\widetilde{V}_{s,\jmath}x_{s,\jmath}[l]
		\label{prob:revised_p1_obj}\\
		\mathrm{s.t.}\quad
		&
		\sum_{a\in\mathcal{A}_s[l]}x_{s,\jmath}[l]=1, \quad
		s\in\mathcal{S}_M^{\mathrm{new}}[l],
		\label{prob:one_profile}\\
		&
		\bm{b}^M(\bm{x};l)
	\preceq
	\underline{\bm{B}}^{\mathrm{av}}[l],
		\label{prob:physical_budget}\\
		&
		\boldsymbol{b}^{M}(\boldsymbol{x};l)
		\preceq
		\boldsymbol{B}_{c_l}^{M},
		\label{prob:coexistence_budget}\\
		&\sum_{(s,\jmath)\in H}\overline x_{s,\jmath}[l]\le|H|-1,\quad H\in\mathcal H_l,
		\label{prob:incompatibility}\\
		&
		x_{s,\jmath}[l]\in\{0,1\}.
		\label{prob:binary}
	\end{align}
\end{subequations}
where $\underline{\bm{B}}^{\mathrm{av}}[l]$ is the component-wise guaranteed lower envelope over the next finite operational window. 

Problem $\textbf{P1}[l]$ is a finite 0-1 linear program because all candidate-dependent capability, risk, delay, and resource terms
are precomputed constants. Constraint Eq.~\eqref{prob:physical_budget} enforces instantaneous physical-resource availability, whereas Eq.~\eqref{prob:coexistence_budget} restricts the
total MCX contract to the calibration-safe inner region for non-MCX
services. Previously admitted mission-horizon contracts remain fixed
and appear through $\boldsymbol{b}^{\mathrm{lock}}[l]$.
Algorithm~1 summarizes the actuation-aware FH-EC MCX profile orchestration in O-RAN.
The proposed orchestration separates capability generation from
real-time decision making. FH-EC estimation and certificate
construction are performed offline, while the online Near-RT RIC only solves a finite-dimensional profile selection problem. Therefore, the real-time complexity is determined by the number of executable profiles rather than the instantaneous radio state space.

\subsection{Simultaneous Post-Selection Validity of	Finite-Library O-RAN Service Contracts}
\textit{Definition 4 (Calibration-compatible O-RAN execution):}
For a feasible profile-selection vector $\bm{x}$ returned by
$\mathbf{P1}[l]$, let
\begin{equation}
	\jmath_s(\bm{x},l)
	\triangleq
	\left(
	a_s,
	c_l,
	e_{s,a_s}(\overline{\bm{x}},c_l),
	\theta_s,
	\bm{q}_s
	\right)
\end{equation}
denote the runtime calibration key of selected nonzero profile
$a_s$. The execution is calibration compatible if:

1) $\jmath_s(\bm{x},l)\in\mathcal J$ for every selected
nonzero profile;

2) The stored O-CU/O-DU policy, participating radio legs, resource contract, and profile identifier are preserved over the complete mission window, and the realized observation age, actuation delay, load, interference, and infrastructure state remain within the corresponding calibrated execution-class envelope;

3) Conditioned on calibration key $\jmath$, the complete
calibration trajectories $\{W_{\jmath}^{(n)}\}_{n=1}^{N_{\jmath}}$
and an independent future mission trajectory $W_{\jmath}^{\rm run}$ are identically distributed, and the 
calibration trajectories are mutually independent.

If any runtime invariant in condition 2) is violated, the
corresponding statistical certificate is declared invalid and
the profile is rejected or transferred to a separately
calibrated fallback class.

\textit{Theorem 1 (Simultaneous post-selection validity of
	finite-library O-RAN service contracts):} Consider any post-calibration orchestration epoch $l$. Suppose that:

1) The E2-context map, finite execution-class partition,
candidate-policy library, service requirements, profile
screening rules, residual non-MCX policy $\psi_N$, and all
hyperparameters are fixed before $\mathcal D^{\rm cal}$ is accessed;

2) The calibration-runtime exchangeability condition in
Definition 4 holds for every calibration key
$\jmath\in\mathcal J$;

3) The confidence-failure budgets satisfy:
\begin{equation}
	\sum_{\jmath\in\mathcal J}
	\left(
	\alpha_{\jmath}^{C}
	+
	\sum_{q\in\mathcal Q_{\jmath}^{+}}
	\alpha_{\jmath,q}
	\right)
	\le
	\alpha_{\rm prof},
\end{equation}
and the non-MCX budgets satisfy the simultaneous condition;

4) $\mathbf{P1}[l]$ enforces the physical-resource constraint,
the calibration-safe non-MCX budget, profile locking, and the
forbidden-hyperedge constraints, and returns only calibration-compatible profile combinations.\\
With calibration confidence at least
$(1-\alpha_{\rm prof}-\alpha_N)$, every selected nonzero profile
$a_s$ simultaneously satisfies:
\begin{align}
	\begin{cases}
	C_{\jmath_s(\bm{x},l)}^{\rm FH} \!\!\!\!\!\!
	&\ge
	\underline C_{\jmath_s(\bm{x},l)}^{\rm FH};
	\\
	p_{\jmath_s(\bm{x},l)}^{q,\rm FH}\!\!\!\!\!\!
	&\le
	\overline p_{\jmath_s(\bm{x},l)}^{q,\rm FH}
	\le
	\epsilon_s^q,
	\qquad
	q\in\mathcal Q_s^{+};
	\\
	\bm{b}^M(\bm{x};l)\!\!\!\!\!\!
	&\preceq
	\bm{b}^{\rm av}[l];
	\\
	U_{N,c_l}^{\psi_N}
	\!\left(
	\bm{b}^M(\bm{x};l)
	\right)\!\!\!\!\!\!
	&\ge
	(1-\rho^{\max})U_{N,c_l}^{0};
	\\
	r_{n,c_l}^{\psi_N}
	\!\left(
	\bm{b}^M(\bm{x};l)
	\right)\!\!\!\!\!\!
	&\ge
	r_n^{N},
	\qquad
	n\in\mathcal S_N^{\rm prot}.
		\end{cases}
	\label{eq:theorem-protected-rate}
\end{align}
The conditional joint multi-QoS mission risk of each
selected MCX flow is bounded as follows:
\begin{equation}
	\Pr_{\rm run}
	\left\{
	\mathcal V_s^{\rm MQ}
	\mid
	c_{l_s}=c_l,\,
	e_{s,a_s}=e_{s,a_s}(\overline{\bm{x}},c_l)
	\right\}
	\le
	\epsilon_s^R
	+
	\sum_{q\in\mathcal Q_s}
	\epsilon_s^q.
	\label{eq:joint-mission-risk-new}
\end{equation}
A prescribed joint mission-risk target $\epsilon_s^{\rm MQ}$ is satisfied whenever the following inequality holds:
\begin{equation}
	\epsilon_s^R
	+
	\sum_{q\in\mathcal Q_s}
	\epsilon_s^q
	\le
	\epsilon_s^{\rm MQ}.
	\label{eq:mission-risk-budget-new}
\end{equation}

\begin{IEEEproof}
	The proof of Theorem~1 is in Appendix B.
\end{IEEEproof}

\begin{algorithm}[t]
	\caption{Actuation-Aware FH-EC MCX Profile Orchestration}
	\label{alg:revised_orchestration}
	\begin{algorithmic}[1]
		\STATE \textbf{Offline profile construction:}
		\STATE Fix the E2-context partition, finite execution-class map, and E2-executable policy library.
		\STATE Construct pre- and post-actuation service kernels.
		\STATE Compute the actuation-aware FH-EC in
		Eq.~\eqref{eq:matrix_actuation_fhec}.
		\STATE Obtain simultaneous capability LCBs and operational-risk UCBs.
		\STATE Construct $\bm{b}_c^M$, the calibrated execution-class library, and the forbidden-hyperedge sets $\{\mathcal H_l\}$.
		\STATE \textbf{Online admission and orchestration at epoch $l$:}
		\STATE Receive the latest E2 information
		$\mathcal{F}_{l}^{\mathrm{E2}}$.
		\STATE Identify the calibrated context cell $c_l$.
		\STATE Retain the locked profiles of
		$\mathcal{S}_M^{\mathrm{lock}}[l]$.
		\STATE Retrieve, rather than re-estimate, the confidence-screened
		profile sets for $\mathcal{S}_M^{\mathrm{new}}[l]$.
		\STATE Solve $\mathbf{P1}[l]$ in
		Eq.~\eqref{prob:revised_p1}.
		\STATE Issue the E2 control actions associated with the selected profiles.
		\STATE Verify the induced execution-class label and monitor the stored policy, resource, load, interference, and actuation-delay envelopes throughout the mission window.
		\STATE After actuation, lock each admitted profile until its mission
		window expires.
		\STATE Release its reserved resources at the end of the mission window.
	\end{algorithmic}
\end{algorithm}

	\section{Performance Evaluations}\label{sec:results}

\subsection{Simulation Setup and Baselines}

We evaluate the proposed FH-EC-driven MCX orchestration framework
in a shared O-RAN scenario. The network simultaneously supports mission-critical MCX services and ordinary mobile services over shared radio and transport resources.
The wireless service process is modeled as a finite-state
Markov-additive process. The service state captures both temporal
channel variations and correlated connectivity failures among
duplicated transmission legs.
Unless otherwise specified, the major simulation parameters are listed in Table~I.

\begin{table}[t]
	\centering
	\caption{Simulation Parameters}
	\label{tab:sim_parameters}
	\begin{tabular}{c|c}
		\hline
		Parameter & Value\\
		\hline
		Logical E2 domains & 4\\
		O-DUs per domain & 3\\
		Slot duration $T_{\rm sl}$ & 1 ms\\
		Mission horizon $\Delta_s$ & 0.1--5 s\\
		QoS exponent $\theta$ & $10^{-3}$--$10^{-1}$\\
		PHY BLER & $10^{-5}$\\
		PDCP duplication legs & 2\\
		E2 observation age & 1--500 ms\\
		Actuation-delay bound/range & 1--20 slots\\
		SC/DP legs &  $1/2$ \\
		\hline
	\end{tabular}
\end{table}

The finite-horizon capability calculation, confidence-calibrated profile library, non-MCX-safe budget, and mixed-integer profile selector are implemented in MATLAB.  Packet generation, queueing, link adaptation, HARQ, infrastructure impairment, and end-to-end packet delivery are implemented in ns-3 with 5G-LENA~\cite{Patriciello2019E2E,Koutlia2022Calibration,11028899}. 5G-LENA provides an end-to-end NR system-level simulation framework and has been calibrated against 3GPP reference scenarios. 


Three E2 exposure configurations are considered. The rich configuration exposes per-flow or fine-grained load, channel-quality, queue, delivery, and node-availability information together with the broadest simulated action set. The moderate configuration exposes aggregated measurements and a reduced set of resource and association actions. The sparse configuration exposes only cell-level measurements and coarse association control.

The following schemes use the same radio resources and packet-level simulator.
\begin{itemize}
	\item \textbf{Static-QPP}: A fixed MCX reservation and strict-priority scheduler, with its reservation tuned on the training set.
	\item \textbf{LT-EC}: The same finite candidate policies and test-time selector, but the finite-horizon capability coordinate is replaced by a stationary long-term EC estimate.
	\item \textbf{FH-SC}: The proposed finite-horizon profile construction and orchestration with DP disabled. 
	\item \textbf{SNC-Delay}: An ORANUS-inspired delay-tail provisioning scheme using the same radio deployment and traffic traces, but without explicit jitter, AoI, and terminal-error constraints~\cite{10621356}.
	\item \textbf{Proposed E2-FH}: The complete E2-conditioned finite-profile framework with adaptive SC/DP selection and the calibration-safe non-MCX resource budget.
	\item \textbf{Full-State Oracle}: A reference controller with current full-state information, zero observation and actuation delay, and access to complete simulated candidate set. 
\end{itemize}




\subsection{Results and Discussion}
\begin{figure}[!t]
	\centering
	\includegraphics[scale=0.44]{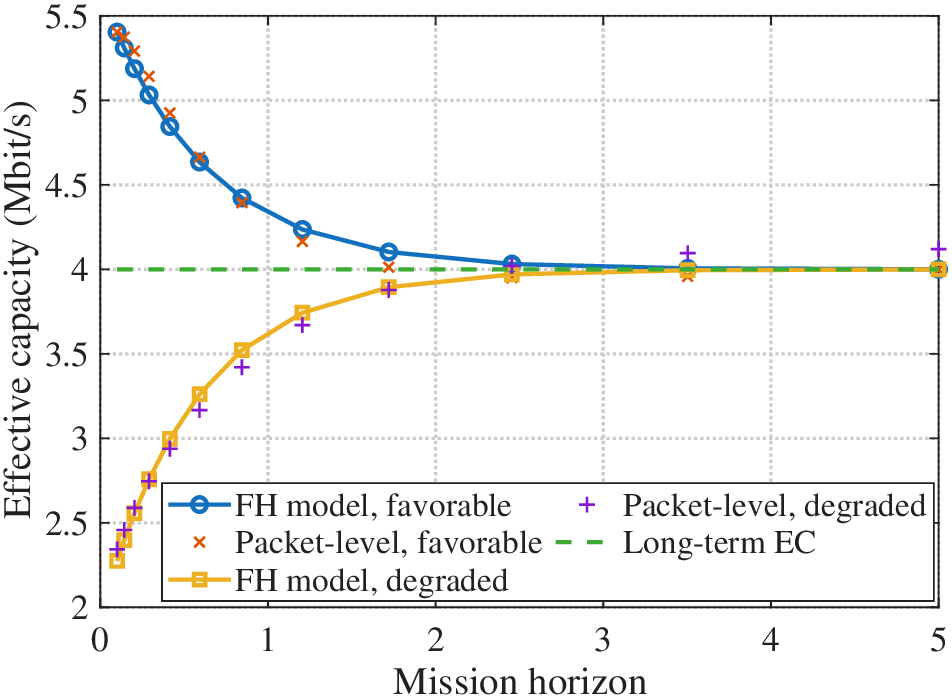}
	\caption{Finite-horizon capability vs. mission horizon $\Delta_{s}$ (s).}
	\label{fig5}
\end{figure}

For each mission horizon $\Delta_s$, independent service trajectories are generated according to the Markov-additive service model. 
Figure~\ref{fig5} compares the analytical finite-horizon capability in Eq.~\eqref{equation038}, with the direct packet-level estimation and conventional long-term effective capacity as functions of the mission horizon $\Delta_{s}$ under favorable and degraded initial E2-conditioned states. The stationary long-term EC is included as the reference. The model follows the packet-level exponential service transform over the tested horizons. 
We can observe from Fig.~\ref{fig5} that at short horizon $\Delta_{s}$=1, the favorable and degraded curves differ by 
about 0.88 Mbit/s, confirming that the initial
E2-conditioned state cannot be replaced by a stationary average.
As shown in Fig.~\ref{fig5}, as the horizon increases, the relative gap between finite-horizon and conventional EC decreases, providing numerical support for Proposition 1.
Fig.~\ref{fig5} also shows that FH-EC captures the transient service capability within short mission windows, while gradually converging to the conventional EC when the horizon becomes sufficiently long.

\begin{figure}[!t]
	\centering
	\includegraphics[scale=0.53]{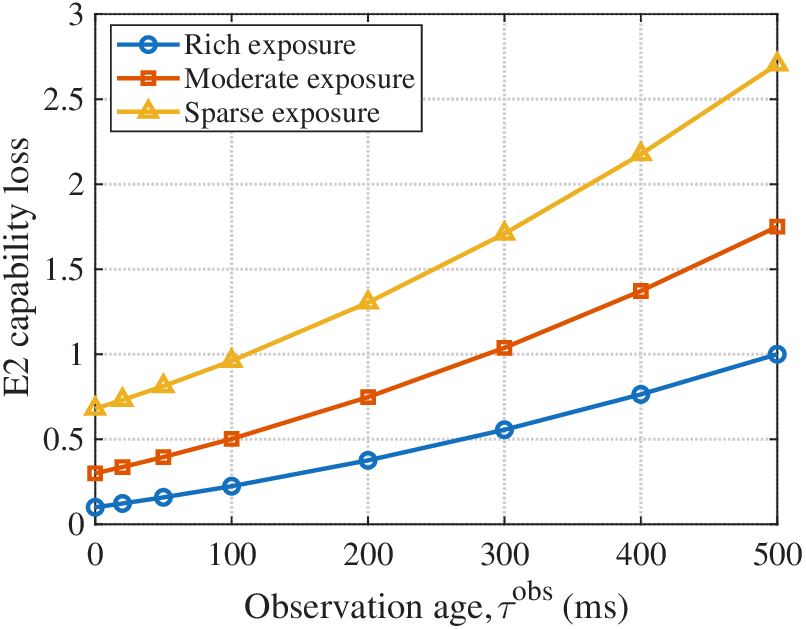}
	\caption{E2 capability loss vs. observation age.}
	\label{fig:e2impact}
\end{figure}

O-RAN introduces additional uncertainty through E2 observation latency and control actuation delay. Fig.~\ref{fig:e2impact} evaluates the impact of E2 exposure on the finite-horizon capability as the observation age $\tau^{\text{obs}}$ varies. 
We can observe from Fig.~\ref{fig:e2impact} that the E2 capability loss increases as the observation age increases.
As shown in Fig.~\ref{fig:e2impact}, rich exposure produces the smallest loss because it retains the most informative observation map and the largest executable action set.
We can observe from Fig.~\ref{fig:e2impact} that outdated E2 information reduces the available MCX capability because the selected profile may no longer match the current network state.

\begin{figure}[!t]
	\centering
	\includegraphics[scale=0.435]{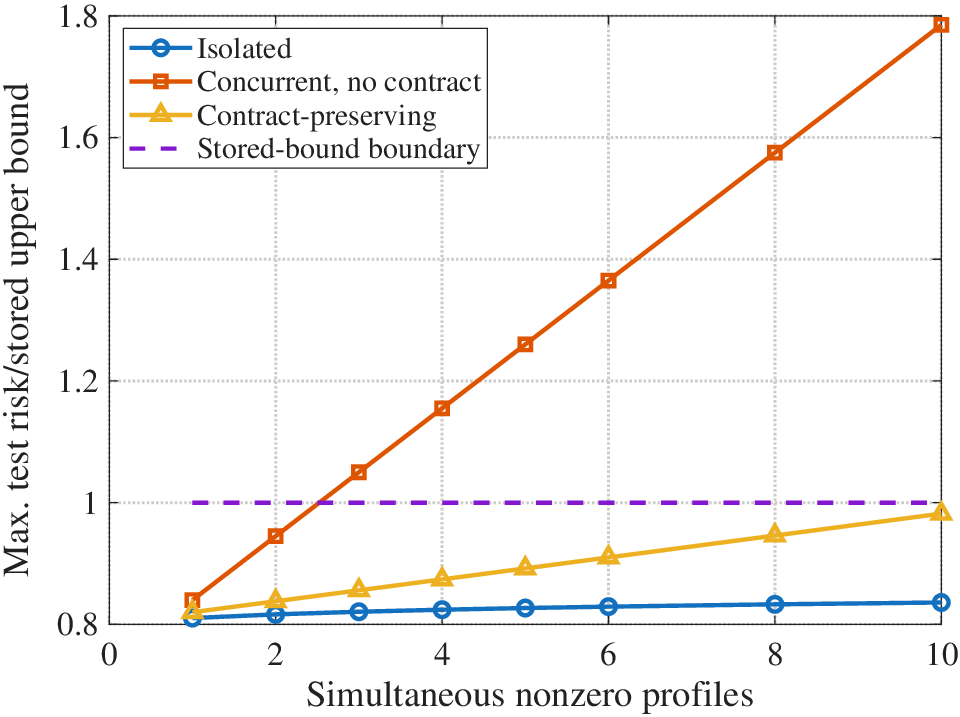}
	\caption{Largest one-sided test upper confidence limit vs. stored profile upper bound.}
	\label{fig8}
\end{figure}

Figure~\ref{fig8} plots the largest one-sided test upper confidence limit among its active operational risks over the corresponding stored profile upper bound to verify the contract-preserving composability assumption. This ratio implies a bound-validity diagnostic. The isolated execution first checks each profile individually. Concurrent execution without enforcing the characterized contracts exceeds the stored bound when more profiles are active.
Under contract-preserving execution, the ratio remains no larger than one. Any profile combination that still exceeds its stored bound is excluded by an incompatibility constraint or recalibrated as a joint profile tuple.

\begin{figure}[!t]
	\centering
	\includegraphics[scale=0.44]{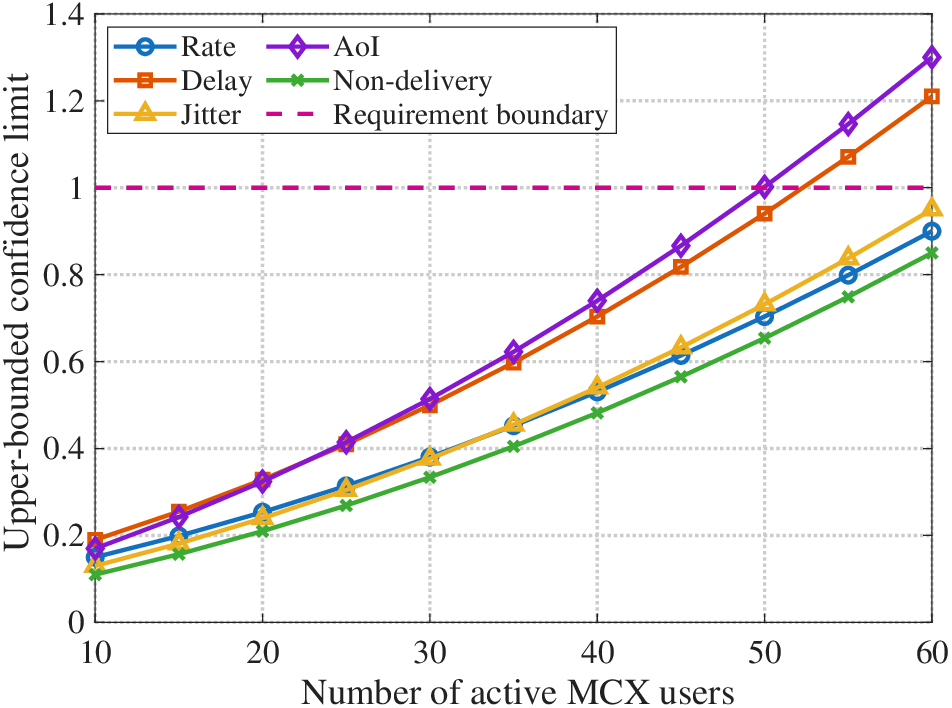}
	\caption{Normalized one-sided risk upper confidence bound vs. number of active MCX users.}
	\label{fig6}
\end{figure}

Figure~\ref{fig6} plots the normalized one-sided risk upper confidence limit of each active delivered-rate, delay, jitter, AoI, and terminal non-delivery risk divided by its requirement with varying numbers of active MCX users. A value no larger than one means that the corresponding pre-existing requirement is satisfied.
We can observe from Fig.~\ref{fig6} that the proposed method satisfies all active dimensions up to 50 simultaneous MCX users. 
It verifies that the improvement is not obtained by reducing delay while silently violating AoI, jitter, or terminal non-delivery requirements. 

\begin{figure}[!t]
	\centering
	\includegraphics[scale=0.495]{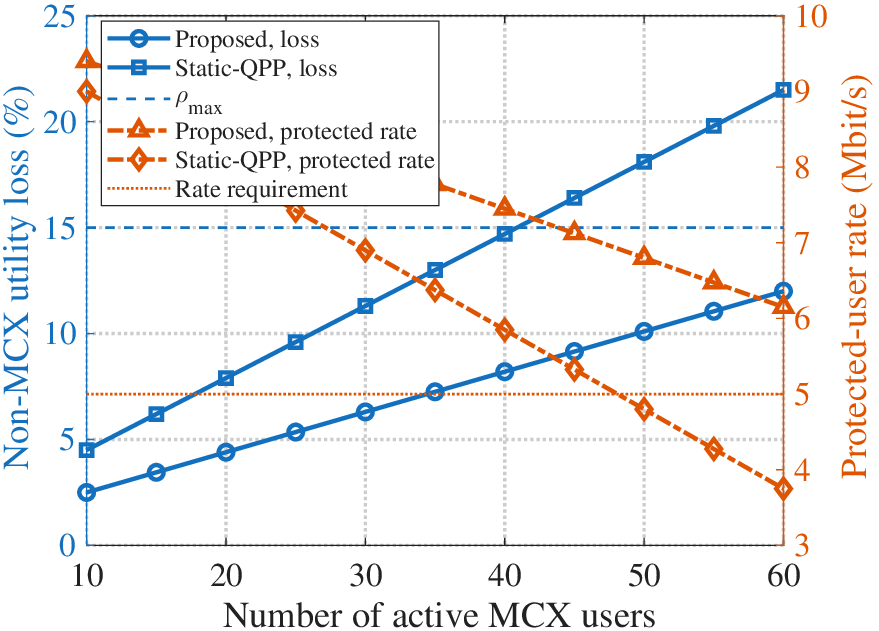}
	\caption{Non-MCX utility loss and protected-user rate vs. number of active MCX users.}
	\label{fig7}
\end{figure}

\begin{figure}[!t]
	\centering
	\includegraphics[scale=0.49]{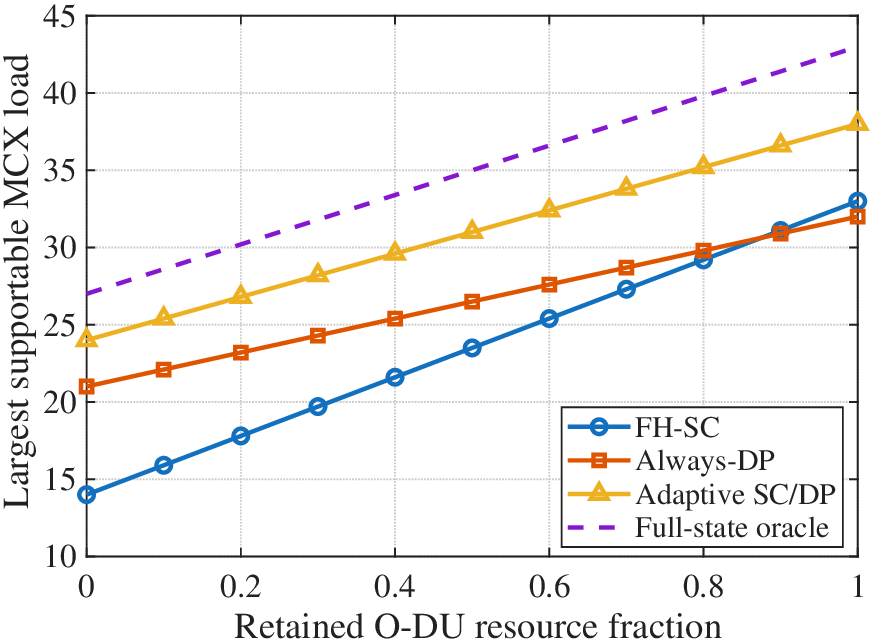}
	\caption{Maximum admissible MCX load vs. the retained resource fraction of the degraded O-DU.}
	\label{fig9}
\end{figure}

Figure~\ref{fig7} presents the corresponding non-MCX utility loss and the delivered rate of protected ordinary users. At the proposed method's largest supportable MCX load, the utility loss is below $\rho_{\max}$. 
Static-QPP either violates the protection conditions at high MCX load or reserves excess resources at low load, whereas the proposed method adapts the selected contracts within $\overline{\bm{b}}_{c_l}^{\mathrm M}$.
This validates the safe-budget construction in Eq.~\eqref{eq:origin_anchored_safe_budget_revised} and Theorem 1.
In addition, Fig.~\ref{fig9} plots the largest supportable admissible MCX load as one O-DU is progressively degraded while comparing with SC-only, always-DP, and the adaptive SC/DP selection. 
We can observe from Fig.~\ref{fig9} that always-DP improves delivery diversity under severe independent degradation but admits fewer flows under normal operation because every copy consumes radio resources. 

	\section{Conclusions}\label{sec:conclusion}
	
We developed an E2-conditioned finite-horizon effective capacity orchestration framework that provides a unified capability 	abstraction for executable O-RAN configurations.
Particularly, we first derived a correlation-aware Markov-additive service model to characterize heterogeneous connectivity options. 
We then proposed the FH-EC expression under finite-blocklength transmission. 	
The resulting FH-EC capability was transformed into confidence-calibrated service profiles, enabling trustworthy profile selection and realization through the Near-RT RIC.
Based on the FH-EC capability abstraction, we developed an 	orchestration framework incorporating contract certification, 	shared-resource protection, and capability-aware profile selection.
Coupled MATLAB/ns-3 evaluations supported the analytical finite-horizon characterization and demonstrated the effects of E2 observation age, actuation delay, concurrent profile execution, and O-DU degradation.

\begin{appendices}
		\section{Proof of Proposition 1}

For notational simplicity, we define
\begin{align}
	\begin{cases}
	d &\triangleq d_{s,a}^{\mathrm{ctl}};\\
	\mathbf{H}
	&\triangleq
	\mathbf{H}_{s,a}^{\mathrm{pre}}(\theta_s);\\
	\mathbf{G}
	&\triangleq
	\mathbf{G}_{s,a}^{\infty}(\theta_s);\\
	\boldsymbol{\alpha}
	&\triangleq
	\boldsymbol{\alpha}_{s,a}[l_s].
		\end{cases}
\end{align}
Using Eq.~\eqref{eq:matrix_actuation_fhec}, we have
\begin{equation}
	C_{s,a}^{\mathrm{FH}}(\theta_s)
	=
	-\frac{1}{\theta_sK_sT_{\mathrm{sl}}}
	\log
	\left[
	\boldsymbol{\alpha}^{T}
	\mathbf{H}
	\mathbf{G}^{K_s-d}
	\mathbf{1}
	\right].
	\label{eq:appendix_fhec}
\end{equation}
Since $\mathbf{G}$ is primitive, the Perron--Frobenius theorem
guarantees a positive spectral radius $\rho$, positive right and
left Perron eigenvectors $\mathbf{v}$ and $\mathbf{u}$, and the
normalization $\mathbf{u}^{T}\mathbf{v}=1$. Moreover, we have
\begin{equation}
	\mathbf{G}^{n}
	=
	\rho^{n}\mathbf{v}\mathbf{u}^{T}
	+
	o(\rho^{n}),
	\qquad n\rightarrow\infty.
	\label{eq:pf_expansion}
\end{equation}
Therefore, we can obtain
\begin{align}
	&\boldsymbol{\alpha}^{T}
	\mathbf{H}
	\mathbf{G}^{K_s-d}
	\mathbf{1}=
	\rho^{K_s-d}
	\left(
	\boldsymbol{\alpha}^{T}\mathbf{H}\mathbf{v}
	\right)
	\left(
	\mathbf{u}^{T}\mathbf{1}
	\right)
	\left[1+o(1)\right].
	\label{eq:pf_scalar_expansion}
\end{align}
The coefficient in Eq.~\eqref{eq:pf_scalar_expansion} is strictly
positive because of $\mathbf{u}^{T}\mathbf{1}>0$.
By taking the logarithm, we have
\begin{align}
	&\log
	\left[
	\boldsymbol{\alpha}^{T}
	\mathbf{H}
	\mathbf{G}^{K_s-d}
	\mathbf{1}
	\right]
	\nonumber\\
	&\quad=
	(K_s-d)\log\rho
	+
	\log
	\left[
	\left(
	\boldsymbol{\alpha}^{T}\mathbf{H}\mathbf{v}
	\right)
	\left(
	\mathbf{u}^{T}\mathbf{1}
	\right)
	\right]
	+o(1).
	\label{eq:log_pf_expansion}
\end{align}
Substituting Eq.~\eqref{eq:log_pf_expansion} into
Eq.~\eqref{eq:appendix_fhec}, using
$d/K_s\rightarrow0$, and letting $K_s\rightarrow\infty$, we obtain
\begin{equation}
	\lim_{K_s\rightarrow\infty}
	C_{s,a}^{\mathrm{FH}}(\theta_s)
	=
	-\frac{1}{\theta_sT_{\mathrm{sl}}}\log\rho,
\end{equation}
which completes the proof of Proposition~1.		

\section*{Appendix B\\Proof of Theorem~1}
Define the simultaneous calibration event $\mathcal E
	\triangleq	\mathcal E_{\rm prof}\cap\mathcal E_N$.
From the simultaneous profile-calibration result, we have $\Pr_{\rm cal}\left\{\mathcal E_{\rm prof}\right\}\ge1-\alpha_{\rm prof}$,
and $\Pr_{\rm cal}\left\{\mathcal E_N\right\}\ge1-\alpha_N$.
By the union bound, we obtain
\begin{equation}
	\Pr_{\rm cal}
	\left\{
	\mathcal E
	\right\}
	\ge
	1-\alpha_{\rm prof}-\alpha_N.
	\label{eq:joint-calibration-event}
\end{equation}
On $\mathcal E_{\rm prof}$, the capability lower bound and all
operational-risk upper bounds hold simultaneously for every
calibration key $\jmath\in\mathcal J$. By Definition 4 and the
forbidden-hyperedge constraints, every nonzero profile selected
by $\mathbf{P1}[l]$ induces a calibration key
$\jmath_s(\bm{x},l)\in\mathcal J$. Consequently, we have
\begin{equation}
	C_{\jmath_s(\bm{x},l)}^{\rm FH}
	\ge
	\underline C_{\jmath_s(\bm{x},l)}^{\rm FH},
\end{equation}
and
\begin{equation}
	p_{\jmath_s(\bm{x},l)}^{q,\rm FH}
	\le
	\overline p_{\jmath_s(\bm{x},l)}^{q,\rm FH},
	\qquad
	q\in\mathcal Q_s^{+}.
\end{equation}
Because these inequalities hold simultaneously over the entire
finite library, selecting the calibration key after observing
the current E2-context cell and solving $\mathbf{P1}[l]$ does
not introduce an additional post-selection confidence penalty.
The profile-screening conditions then imply
\begin{equation}
	\overline p_{\jmath_s(\bm{x},l)}^{q,\rm FH}
	\le
	\epsilon_s^q,
	\qquad
	q\in\mathcal Q_s^{+}.
\end{equation}
Constraint in Eq.~\eqref{prob:physical_budget} directly proves
$\bm{b}^M(\bm{x};l)
	\preceq
	\bm{b}^{\rm av}[l]$.
On $\mathcal E_N$, the simultaneous non-MCX rate bounds, the
definition of the calibration-safe set, and constraint Eq.~\eqref{prob:coexistence_budget}  give
\begin{equation}
	U_{N,c_l}^{\psi_N}
	\left(
	\bm{b}^M(\bm{x};l)
	\right)
	\ge
	(1-\rho^{\max})U_{N,c_l}^{0},
\end{equation}
and
\begin{equation}
	r_{n,c_l}^{\psi_N}
	\left(
	\bm{b}^M(\bm{x};l)
	\right)
	\ge
	r_n^{N,\rm prot},
	\qquad
	n\in\mathcal S_N^{\rm prot}.
\end{equation}
Finally, by applying the union bound to the delivered-rate and
active dimension-specific operational violation events, we have
\begin{align}
	&\Pr_{\rm run}
	\left\{
	\mathcal V_s^{\rm MQ}
	\mid
	c_{l_s}=c_l,\,
	e_{s,a_s}=e_{s,a_s}(\overline{\bm{x}},c_l)
	\right\}
	\nonumber\\
	&\qquad\le
	p_{\jmath_s(\bm{x},l)}^{R,\rm FH}
	+
	\sum_{q\in\mathcal Q_s}
	p_{\jmath_s(\bm{x},l)}^{q,\rm FH}
\le	\epsilon_s^R+\sum_{q\in\mathcal Q_s}
	\epsilon_s^q.
\end{align}
Combining this result with Eq.~\eqref{eq:joint-calibration-event} proves Theorem~1.

\end{appendices}

	\nocite{*}
	\footnotesize
	\bibliographystyle{IEEEtran}
	\bibliography{myref.bib}

\end{document}